\documentclass{jfm}
\usepackage{graphicx}
\usepackage{epstopdf, epsfig}
\usepackage{amsfonts,amssymb,amsmath,mathbbol}   

\usepackage{xcolor} 
\usepackage{float} 
\usepackage{float}
\usepackage{placeins}
\usepackage{physics}

\def\Sh{{\it Sh}}
\def\Sc{{\it Sc}}

\shorttitle{Evaporating droplets}
\shortauthor{S. Chakraborty, S. S. Ade, A. J. Tudu, L. D. Chandrala and K. C. Sahu}

\title{Evaporation of a freely floating droplet in an airstream: effects of temperature, humidity, and shape oscillations}

\author{Shubham Chakraborty,\aff{1} Someshwar Sanjay Ade,\aff{1} Aman John Tudu,\aff{2} Lakshmana Dora Chandrala\aff{3} \and Kirti Chandra Sahu\aff{2}\corresp{\email{lchandrala@mae.iith.ac.in, ksahu@che.iith.ac.in}}}

\affiliation{\aff{1} Center for Interdisciplinary Program, Indian Institute of Technology Hyderabad, Sangareddy - 502 284, Telangana, India
\aff{2}Department of Chemical Engineering, Indian Institute of Technology Hyderabad, Sangareddy - 502 284, Telangana, India
\aff{3}Department of Mechanical and Aerospace Engineering, Indian Institute of Technology Hyderabad, Sangareddy - 502 284, Telangana, India}

\begin{document}

\maketitle

\begin{abstract}
We present a comprehensive experimental and theoretical investigation of the evaporation dynamics of freely levitated water droplets in an upward airstream under varying temperature and relative humidity conditions, using a custom-designed wind tunnel that replicates natural rainfall scenarios. A high-speed imaging system captures the temporal evolution of morphology, shape oscillations, and size reduction of the droplet undergoing evaporation. Our observations reveal that larger droplets exhibit persistent shape oscillations due to the interplay between inertia and surface tension in the presence of convective airflow, which significantly alters the evaporation rate compared to that of a stationary spherical droplet in quiescent air. To quantify the effects of air convection, complex morphology, and shape oscillations of the levitated droplet at different temperatures and humidity, we develop a modified evaporation model that extends the classical $d^2$-law. This model incorporates (i) a generalized Sherwood number that accounts for the variation in Reynolds number, Schmidt number, temperature, and relative humidity and (ii) a shape factor that captures the time-averaged surface area of oscillating droplets. The model is validated against experimental findings across a wide range of droplet sizes and environmental conditions, showing excellent agreement in predicting the temporal evolution of droplet diameter and total evaporation time. Furthermore, we construct a regime map showing the variation in the lifetime of the droplet in the temperature-humidity space. The present study establishes a framework that integrates convective transport and morphological deformation, offering new insights into the microphysics of raindrop evaporation. 
\end{abstract}



\section{Introduction} \label{sec:intro}

The evaporation of droplets is a fundamental process encountered in a wide range of industrial applications, such as spray combustion, drying, and cooling technologies \citep{huang1990evaporation,sazhin2006advanced,pinheiro2019evaluation}, biological systems \citep{mittal2020flow} and various natural phenomena \citep{houghton1933study,best1952evaporation,caplan1966evaporation,abraham1962evaporation,tardif2010evaporation}. During rainfall, evaporation alters the shape and size of falling raindrops through continuous heat and mass transfer with the surrounding air. As raindrops descend through the atmosphere, they encounter progressively warmer and drier conditions, leading to complex evaporation dynamics that significantly influence the evolution of raindrop size distributions. In addition to modulating rainfall, droplet evaporation has far-reaching implications for understanding cloud microphysics, quantifying precipitation intensity, and predicting the vertical redistribution of latent heat within the atmosphere. Thus, a detailed understanding of droplet-scale evaporation is essential to improve the accuracy of predictive models related to cloud development, raindrop formation, and atmospheric moisture transport \citep{beard1971wind,beji2018detailed}. The importance of evaporation in influencing both natural and engineering processes has been recognized for more than a century \citep{langmuir1918evaporation,apashev1962evaporation,beard1971wind}, and it remains an active area of research even today \citep{schlottke2008direct,tripathi2015evaporating,pal2023accurate,sezen2023water,pal2024modeling,pal2024oblique}.

A large volume of experimental investigations on evaporation have mainly focused on sessile droplet configurations, examining the influence of substrate wettability and surrounding ambient conditions \citep{birdi1989study,deegan1997capillary,shahidzadeh2006evaporating,sefiane2011expression,gurrala2019evaporation,katre2021evaporation,diddens2021competing}. These studies have also contributed to the development of theoretical models for predicting evaporation rates by considering droplets under constant contact radius (CCR) and constant contact angle (CCA) modes, incorporating mechanisms such as pure diffusion, free convection, and passive vapor transport. Furthermore, several studies have examined the distribution of the vapor concentration, the evaporative flux above the droplet surface, and the influence of Marangoni flows on the evaporation dynamics \citep{picknett1977evaporation,erbil2012evaporation,saenz2017dynamics,masoudi2017axisymmetric}. In addition to sessile droplets, a few studies have also investigated pendant droplet configurations \citep{picknett1977evaporation,erbil2012evaporation,pandey2020dynamics}. Due to their restricted dynamical behaviour, sessile and pendant droplets are relatively easier to investigate experimentally than falling droplets due to the simplicity of the experimental setup, better control over boundary conditions, and their suitability for detailed observation using imaging techniques. 

In addition to sessile and pendant droplet configurations, several studies have investigated the evaporation of acoustically levitated droplets \citep{yarin1998acoustic,yarin1999evaporation,yarin2002evaporation,brenn2007evaporation,saha2010experimental,sasaki2019transition,maruyama2020evaporation}, albite in quiescent environments. Unlike sessile or pendant droplets, which are constrained by solid or supporting surfaces, acoustically levitated droplets are freely suspended in air at the pressure nodes of high-intensity sound waves. This contactless configuration eliminates surface-related effects such as heat conduction, contact-line dynamics, and surface contamination, thereby enabling a more intrinsic investigation of evaporation behaviour. Over the past few decades, numerous researchers have advanced the understanding of acoustically levitated droplet evaporation, progressing from fundamental studies of droplet deformation to detailed analyses of multicomponent and multiphase evaporation. Early theoretical and experimental works by \citet{yarin1998acoustic,yarin1999evaporation,yarin2002drying,yarin2002evaporation} established the influence of acoustic radiation pressure on droplet shape, heat transfer, and mass transfer, revealing that the acoustic field strongly governs evaporation dynamics. Subsequent investigations by \citet{brenn2007evaporation} and \citet{saha2010experimental} incorporated convective effects and laser heating to examine the thermophysical evolution of acoustically levitated droplets. Later studies by \citet{bjelobrk2012acoustic} and \citet{sasaki2019transition} provided further insights into internal circulation and interfacial flow and their roles in evaporation dynamics. More recent research by \citet{maruyama2020evaporation,mitsuno2025evaporation,wakata2024evaporation,krishan2024evaporation} explored multicomponent evaporation and phase transitions through both experimental and numerical approaches. Collectively, these works have significantly deepened the understanding of evaporation under acoustic confinement. However, the dominant influence of the acoustic field often obscures the role of forced convection \citep{yarin1999evaporation}. In contrast, falling droplets such as raindrops exhibit complex evaporation dynamics arising from shape oscillations driven by the interplay of viscous, inertial, and surface-tension forces, together with interactions with the surrounding airstream.

In the context of raindrops, the early works of \cite{beard1971wind,pruppacher1979wind,mitra1992wind} pioneered wind tunnel experiments to investigate the evaporation behaviour of water droplets falling as a spray from a nozzle, subjected to an upward-moving airstream. These studies quantified evaporation rates over a wide range of droplet sizes, from tens of microns to several millimetres, and developed empirical correlations to capture the influence of convective airflow on droplet evaporation. As a droplet falls, it undergoes natural shape oscillations, transitioning between oblate, spherical, and prolate morphologies, due to the competition between inertial and surface tension forces. The work of \cite{pruppacher1971semi} revealed that small droplets ($d < 1$ mm) typically remain spherical due to the dominance of surface tension. In contrast, larger raindrops ($d > 1$ mm) often become non-spherical, assuming oblate or prolate shapes, or exhibiting more complex deformations \citep{tsamopoulos1983nonlinear,beard1989natural,szakall2009wind,szakall2010shapes,szakall2014wind,agrawal2017nonspherical,balla2019shape,balla2020numerical}. Small spherical raindrops attain terminal velocity and a stable equilibrium shape when the upward drag force balances their weight. However, for larger droplets, the relatively weaker surface tension cannot fully counteract inertial deformation, leading to sustained shape oscillations during their descent \citep{balla2019shape,agrawal2017nonspherical,agrawal2020experimental,feng1991perturbation}. The time period ($t_p$) of these natural oscillations for a freely suspended water droplet is given by \citep{rayleigh1879capillary,nelson1972oscillation,howarth1979one,balla2019shape}:
\begin{equation}
t_{p}=2\pi \left[ \frac{(n+1)\rho_w+n\rho_a}{n(n+1)(n-1)(n+2)}  \left ( \frac{d^3}{8\sigma} \right) \right]^{1/2}, \label{Tp_Theory}
\end{equation}
where $\rho_w$ and $\rho_a$ denote the densities of the liquid and the surrounding medium, respectively; $\sigma$ represents the surface tension; $d$ is the equivalent spherical diameter of the droplet; and $n$ is the oscillation mode, with the fundamental mode corresponding to $n = 2$. The frequency ($\omega$) of these natural oscillations is given by $\omega=1/t_p$. These shape oscillations play a crucial role in influencing the evaporation behaviour of raindrops, particularly as they descend from the cold, highly humid environment within clouds to the warmer, less humid layers of the atmosphere.

The classical $d^2$-law provides a framework for modelling the evaporation rate of a stationary spherical droplet in a quiescent environment, assuming a constant evaporation flux \citep{ranz1952evaporation,spalding1960standard,law1982recent}. It is given by
\begin{equation}
d^2 = {d_0}^2 - K t,
\end{equation}
where $d_0$ is the initial droplet diameter, $t$ denotes time and $K$ is the evaporation rate constant. This model assumes a quasi-steady, diffusion-limited evaporation process in which the droplet remains spherical, and the vapour concentration at the liquid–gas interface is in equilibrium. The evaporation rate constant $K$ depends on various factors, including ambient temperature and pressure, vapour diffusivity, latent heat of vaporization, thermal conductivity, and the physical properties of the liquid and the surrounding gas. Subsequently, several extensions of the classical evaporation model have been developed to incorporate the effects of convection and droplet deformation \citep{tonini2014evaporation}. While these models offer reasonable predictions for the evaporation of small droplets at low Reynolds numbers \citep{huang1990evaporation,pal2023accurate}, significant deviations from the classical $d^2$-law are observed under more realistic conditions involving convective flows, temperature gradients, internal recirculation, and droplet shape oscillations. To improve the prediction, recent studies have proposed power-law correlations for the Sherwood number that account not only for Reynolds and Schmidt numbers, but also for the effects of ambient temperature and relative humidity \citep{beji2018detailed,nugraha2022sherwood}. Furthermore, direct numerical simulations (DNS) have been employed to investigate the evaporation of highly deformed droplets in realistic environmental conditions \citep{schlottke2008direct,tripathi2015evaporating,irfan2017front,reutzsch2020consistent,pal2023accurate}. The three-dimensional numerical simulations by \cite{reutzsch2020consistent,schlottke2008direct,tripathi2015evaporating} demonstrated that a deformed droplet falling under gravity develops a fore–aft asymmetric vapor envelope. This asymmetry arises due to vapour transport toward the wake region behind the droplet and intensified evaporation in the associated low-pressure zone, resulting in notable deviations from the predictions of the classical $d^2$-law. Furthermore, \cite{reutzsch2020consistent} examined the evaporation behaviour of an oscillating droplet, emphasizing the influence of phase change on droplet morphology, thereby highlighting the complex coupling between interfacial deformation and evaporation during the free fall of a non-spherical droplet.

Additionally, several advancements in modeling have been made to account for the effects of convection, vapor accumulation, and liquid-phase transport on droplet evaporation. Numerical simulations by \citet{renksizbulut1983experimental} provided a coupled solution of the momentum, energy, and species equations for droplets exposed to high-temperature airflow, demonstrating how local gas properties influence the heat transfer rate at the droplet interface. Subsequently, \citet{whang1997experimental} experimentally investigated the convective ignition of suspended fuel droplets, establishing the dependence of flame propagation and ignition delay on flow conditions, as well as the temporal evolution of droplet size. Building on this understanding, \citet{waheed2002mass} numerically examined mass transfer from droplets under combined free and forced convection. They found that the Sherwood number ($Sh$) and, consequently, the evaporation rate are strongly dependent on diffusion and the surrounding flow field. Since the rates of heat and mass transfer are also influenced by the surface area exposed to the flow, droplet shape plays a crucial role in determining evaporation behavior. \citet{li2014theoretical} derived analytical correlations for the Nusselt number ($Nu$) and Sherwood number ($Sh$) for spheroidal droplets under forced convection, revealing that deformed droplets exhibit enhanced mass transfer compared to spherical ones of equal volume. \citet{zhou2015analytical} combined three-dimensional (3D) simulations and analytical modeling to show that evaporating droplets experience reduced drag compared to non-evaporating ones, thereby modifying the interfacial transport rates. Similarly, \citet{ni2010fuel} developed an improved model that simultaneously accounts for interfacial heat and mass transfer, emphasizing the coupled thermodynamic effects at the vapour–liquid interface. A comprehensive theoretical framework unifying these physical processes was later presented by \citet{sirignano2010droplets} to describe realistic droplet vaporization and combustion phenomena. In their formulation, transient convection, Stefan flow, and multicomponent diffusion were incorporated, which the conventional $d^2$-law fails to capture accurately. This formulation serves as the theoretical foundation for the modeling approach adopted in the present work.

As the brief review indicates, there remains considerable scope for advancing our understanding of evaporating droplets, particularly through experiments involving falling raindrops undergoing shape oscillations under realistic atmospheric conditions that include variations in temperature, humidity, and droplet size. Although some experimental studies have been conducted in the context of rainfall \citep{beard1971wind,pruppacher1979wind,mitra1992wind}, these investigations primarily focused on average evaporation characteristics of sprays consisting of multiple droplets, rather than developing a fundamental understanding of the evaporation dynamics of individual droplets. Furthermore, existing theoretical models of droplet evaporation require refinement to accurately capture the effects of shape oscillations and variations in temperature and humidity. Addressing these gaps is the primary focus of the present experimental and theoretical investigation. In our experiments, a droplet is levitated by an upward airstream set to the terminal velocity of the droplet, with the airflow continuously adjusted through a feedback-controlled system to keep the droplet suspended in the test section as it evaporates. The inlet air is conditioned to maintain specific temperature and relative humidity levels encountered by the levitated droplet, enabling a systematic investigation of the effects of initial droplet size, shape oscillations, ambient temperature, and humidity on the evaporation rate. Based on the experimental findings, we propose an improved theoretical model that accurately predicts the droplet lifetime (i.e., the total evaporation time) under realistic environmental conditions. 

The remainder of this paper is organized as follows. Section~\S\ref{sec:expt} describes the experimental setup and the post-processing methodologies. Section~\S\ref{sec:dis} presents the results, including the dynamics of droplet morphology and shape oscillations under varying temperature and humidity conditions. A theoretical model is developed to predict the evaporation time as a function of droplet size, ambient temperature, and humidity. This section also includes a phase diagram illustrating the droplet lifetime in the temperature–humidity space for typical raindrops. Concluding remarks are provided in Section~\S\ref{sec:conc}.

\section{Experimental setup and procedure} \label{sec:expt}

\subsection{Experimental setup}

\begin{figure}
\centering
\includegraphics[width=0.65\textwidth]{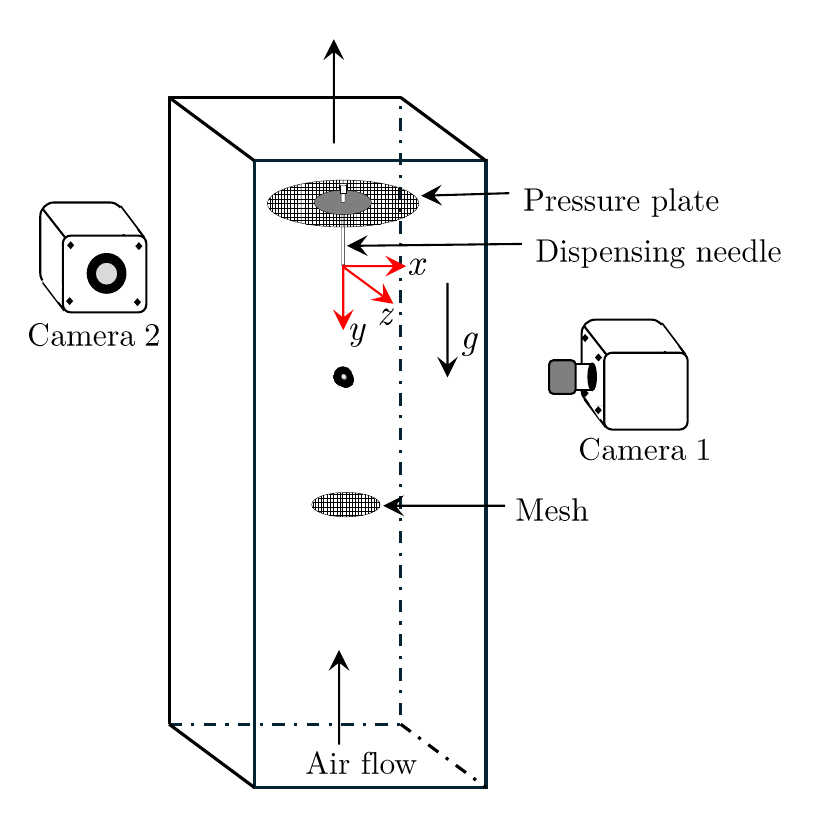}
\caption{A schematic diagram of the experimental test section, where a droplet undergoing evaporation is levitated between a pressure plate at the top and a mesh at the bottom. The droplet is introduced using a dispensing needle. A photograph of the actual experimental setup is provided in Figure~\ref{RRF}.}
\label{schematic}
\end{figure}

We experimentally investigate the evaporation dynamics of a distilled water droplet levitated in an upward airstream using the shadowgraph technique. A schematic diagram of the experimental test section, fabricated from a 10 mm thick transparent acrylic sheet and having internal dimensions of $10~\mathrm{cm} \times 10~\mathrm{cm} \times 45~\mathrm{cm}$, is shown in figure~\ref{schematic}. The setup comprises a dispensing needle connected to a syringe pump for droplet injection, a pressure plate at the top, and a mesh with approximately 10 wires per centimetre at the bottom to enable droplet levitation. The wire mesh generates a localized velocity well in the flow, allowing a droplet to remain suspended near the centre of the test section without drifting toward the walls due to turbulence \citep{Kamra1991}. To ensure the stability of this velocity well, the pressure plate and mesh, separated by 150 mm, are supported by extremely thin threads anchored to the acrylic walls, minimizing interference with the airflow inside the test section. The pressure plate introduces back pressure into the airstream and deflects the flow outward, thereby preventing the collapse of the velocity well \citep{blanchard1950behavior}. In the absence of this plate, the vertical velocity at the centre of the airstream increases, causing the droplet to accelerate upward and lose its stable levitated position. Notably, the pressure plate has a diameter three times larger than that of the mesh (which is approximately 1 cm), significantly enhancing its effectiveness in developing the flow and sustaining the velocity well. This configuration was determined after several trials to ensure optimal levitation conditions. 

\begin{figure}
\centering
\includegraphics[width=0.95\textwidth]{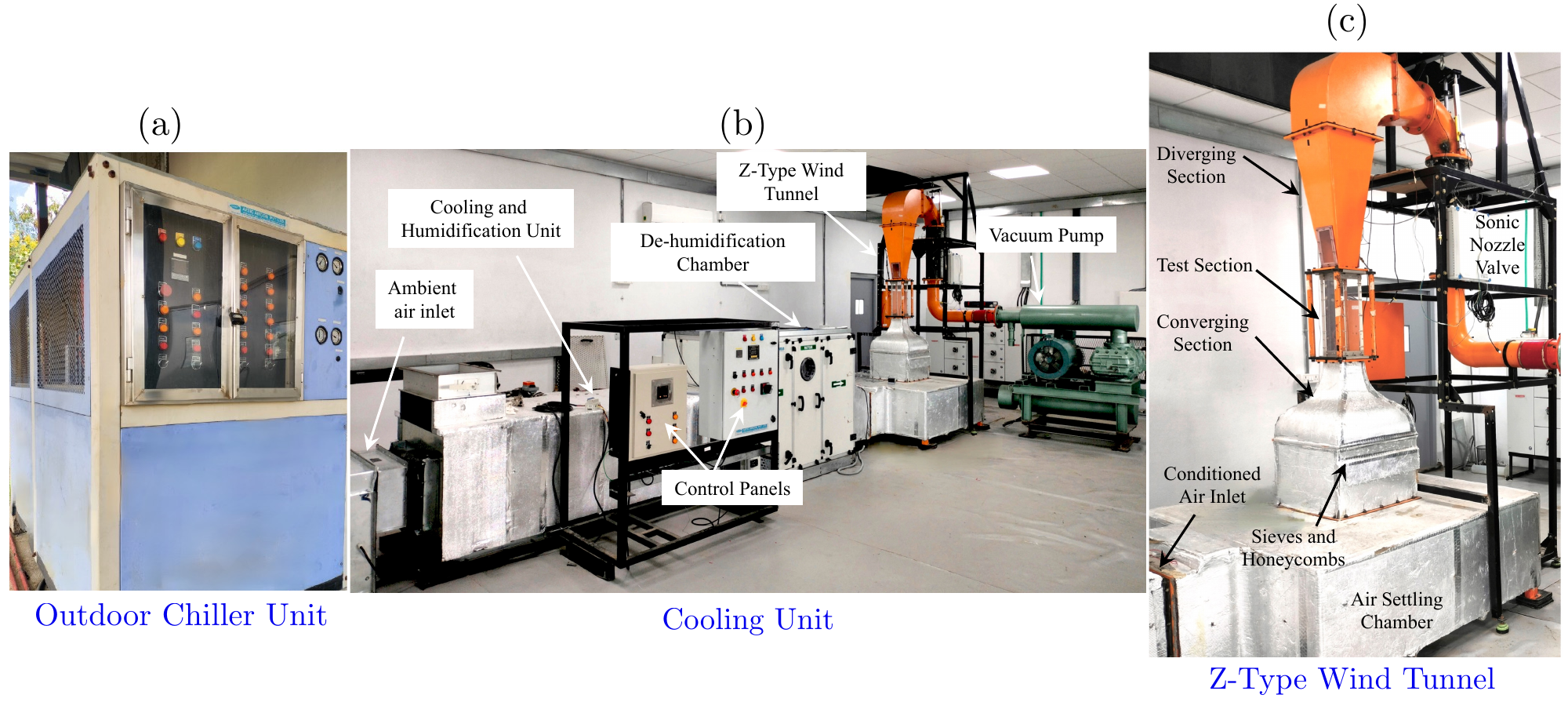}
\caption{A photograph of the state-of-the-art raindrop research facility, designed to simulate a wide range of atmospheric conditions by precisely controlling the temperature and relative humidity of the inlet air \citep{patent_RRF}. It consists of: (a) an outdoor chiller unit, (b) a cooling unit comprising an air inlet section, cooling and humidification unit, dehumidification chamber, and control unit, and (c) a Z-type wind tunnel composed of a settling chamber, a converging dome with sieves and honeycombs, a test section, a diverging section, a sonic nozzle valve, and a vacuum pump.}
\label{RRF}
\end{figure}

The test section is part of a state-of-the-art raindrop research facility (figure~\ref{RRF}), designed to simulate a wide range of atmospheric conditions by controlling the temperature and relative humidity of the inlet air \citep{patent_RRF}. The facility consists of a Z-type vertical wind tunnel integrated with primary and secondary air-conditioning systems, a vacuum pump, and several auxiliary components, including:
(i) a dryer and humidifier,
(ii) a particle filter,
(iii) an air-conditioning system with double-feedback loop temperature control,
(iv) a settling chamber,
(v) honeycombs and sieves for flow straightening,
(vi) a converging section, a test section, a diverging section, and a sonic nozzle valve, and
(vii) flow control mechanisms, including vacuum pumps equipped with variable frequency drives (VFDs). The customized twin-lobe vacuum pump (50 HP/1450 rpm) is capable of generating a flow rate of up to $1500~\textrm{m}^3/\textrm{hr}$, thereby enabling average air velocities in the test section of up to $15~\textrm{m/s}$. The ambient air is first filtered and then directed into the air-conditioning section, which comprises two main units: one for cooling and humidification, and another for dehumidification. The cooling and humidification unit employs a double-feedback loop for precise temperature regulation, while a steam injector adds controlled moisture to the airflow. The dehumidification unit features a desiccant wheel and heaters to remove moisture when required effectively. This system enables precise control of air conditions, with temperature ($T$) adjustable between $-10^{\circ}$C and $40^{\circ}$C, and relative humidity ($RH$) ranging from 5\% to 95\%, both regulated simultaneously via an integrated control panel. Using this dual-stage air-conditioning system, the inlet ambient air is conditioned to attain the desired thermal and moisture levels, effectively replicating the atmospheric environment encountered by falling raindrops. The conditioned air enters the Z-type vertical wind tunnel through a settling chamber, which ensures a smooth and uniform transition of flow from the air-conditioning unit into the tunnel. The airflow is subsequently laminarized by a honeycomb structure, followed by a series of sieves. The air then passes through a converging dome, designed based on a seventh-order polynomial profile, to establish a uniform vertical velocity profile across the test section.

The air exiting the test section flows through a diverging section, followed by a sonic valve. The diverging section decelerates the airflow and facilitates pressure recovery. To maintain a stable levitation of an evaporating droplet, whose size continuously decreases, the airflow speed in the test section is dynamically regulated by the sonic valve using an actuator. In the nozzle section of the sonic valve, the flow accelerates and reaches sonic velocity at the throat, which is immediately followed by a downstream sonic barrier. This barrier prevents pressure fluctuations at the vacuum pump inlet from propagating upstream and disturbing the flow within the test section. As the flow at the throat remains sonic, the mass flow rate of air is directly proportional to the cross-sectional area of the throat. Thus, by adjusting this area, the flow rate and air velocity in the test section can be precisely controlled. 

In our experiments, droplets are generated using a syringe pump (model: HO-SPLF-2D; make: Holmarc Opto-Mechatronics Pvt. Ltd, India) connected to the needle via a silicone tube. To facilitate smooth droplet dispensing, a larger glass tube with an internal diameter of 18 mm is positioned around the needle but placed well above the test section, within the diverging section of the experimental setup. Using this configuration, droplets of various initial diameters ($d_0 = 3.0–5.0$ mm) are levitated throughout the evaporation process by precisely controlling the upward velocity of conditioned air through a sonic nozzle valve and a vacuum pump operated via a feedback mechanism. Since evaporation may begin during the detachment of the droplet from the needle, the droplet volume is measured only after it has completely detached. This instant is represented as $\tau = 0$, and the corresponding initial volume of the droplet is denoted as $V_0$. Shadowgraphy is employed to capture the evaporation and oscillation dynamics of the levitated droplets under varying temperature and humidity conditions. In order to capture two orthogonal views of the droplet required for volume estimation, two high-speed cameras (Model: Phantom VEO 640L; Manufacturer: Vision Research, USA) are used, each equipped with a 135 mm Nikkor lens with a minimum aperture of $f/2$. As shown in figure~\ref{schematic}, the cameras are positioned orthogonally on a horizontal plane and synchronized using a digital delay generator. Uniform background illumination is provided by a high-power LED light source (Model: MultiLED QT; Manufacturer: GSVITEC, Germany) coupled with a diffuser sheet. The captured images have a resolution of $640 \times 1600$ pixels, with an exposure time of 5 $\mu$s and a spatial resolution of 64.53 $\mu$m/pixel. To capture the temporal evolution of droplet shape oscillations, images are recorded at a high frame rate of 600 frames per second (fps). In contrast, since evaporation is a slower process, its dynamics are recorded at 24 fps by conducting separate experiments. Image sequences are initially stored in the internal memory of the cameras and subsequently transferred to a computer for post-processing. A Cartesian coordinate system $(x, y, z)$, with its origin located at the exit of the droplet dispensing needle as shown in figure~\ref{schematic}, is employed to analyze the results discussed in the following section. We have performed Particle Image Velocimetry (PIV) measurements to visualize and quantify the velocity field within the wind-tunnel test section, which includes a pressure plate at the top and a mesh at the bottom. A detailed description of the PIV setup and the corresponding velocity field characterization is provided in Appendix \ref{appendix_PIV}. The measured velocity profiles for different flow rates at three vertical locations, namely at $y = 5$ cm, $y = 7$ cm, and $y = 9$ cm from the tip of the dispensing needle, are shown in figure \ref{vel_prof}(a–c), respectively (Appendix \ref{appendix_PIV}).

\subsection{Post-processing} \label{sec:post-processing}

A sequence of image processing steps was carried out to enhance image quality and accurately extract geometric features from each frame. First, a median filter with a kernel size of $5 \times 5$ was applied to reduce noise while preserving edges. The filtered image was then binarized using a fixed intensity threshold value of 70. To ensure correct segmentation, the binary image was complemented so that the objects of interest appeared as white regions on a black background. Next, small unwanted regions with an area less than 10 pixels were removed to eliminate noise. A morphological closing operation was then performed using a disk-shaped structuring element with a radius of 50 pixels. This step ensured that small gaps were closed and fragmented regions were connected. Subsequently, holes within the segmented objects were filled to obtain solid and continuous shapes. Following segmentation, connected component analysis was conducted to identify individual objects in each frame. For each detected object, geometric properties were extracted using the regionprops function in \textsc{Matlab}$^{\circledR}$ software. These features were used for further quantitative analysis of the droplet shape undergoing evaporation. For each set of parameters, experiments were performed three times, and the results are presented with error bars representing the standard deviation of these repetitions. 

\begin{figure}
\centering
\includegraphics[width=0.90\textwidth]{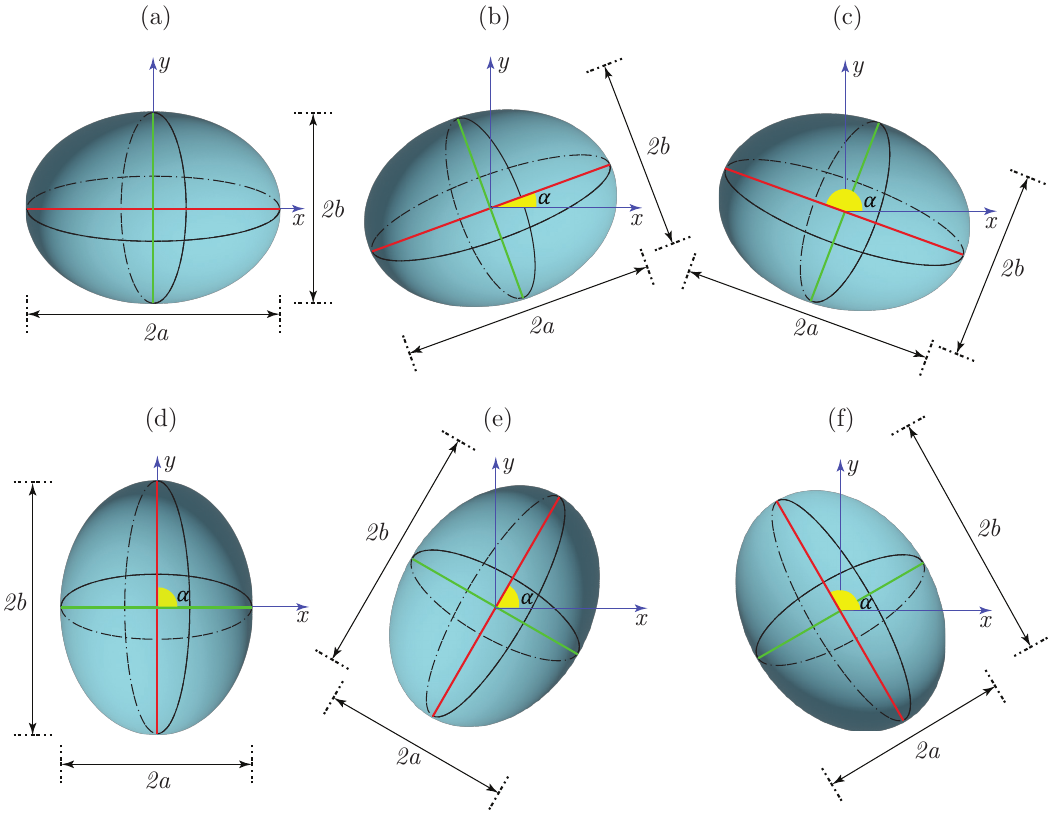}
\caption{Illustration of the procedure for selecting major and minor diameters of non-spherical droplets in various orientations to estimate their volume ($V$) at different time instants. Here, $\alpha$ denotes the orientation angle of the major axis relative to the horizontal. Panels (a) $\alpha = 0^\circ$, (b) $0^\circ \leq \alpha \leq 45^\circ$, and (c) $135^\circ \leq \alpha \leq 180^\circ$ correspond to oblate droplets in different orientations. Similarly, panels (d) $\alpha = 90^\circ$, (e) $45^\circ \leq \alpha \leq 90^\circ$, and (f) $90^\circ \leq \alpha \leq 135^\circ$ represent prolate droplets in various configurations.}
\label{fig:orientation}
\end{figure}

To estimate the droplet volume and to evaluate its geometric properties, data from two orthogonal views were analyzed. Geometric parameters of the droplets, including axis lengths and orientation angles ($\alpha$), were obtained from the processed image data. Since droplets can appear either prolate (elongated) or oblate (flattened) depending on their orientation, adjustments were made to the orientation angle data to correctly account for symmetry. Figure~\ref{fig:orientation} illustrates the procedure for selecting the major and minor diameters of a non-spherical droplet in various orientations. A similar approach has been adopted by \cite{agrawal2020experimental}. For cases where $\alpha < 90^\circ$, $180^\circ$ was added to reflect symmetry about the corresponding axis. The volume of each droplet was then calculated using the formula for a triaxial ellipsoid:
\begin{equation} 
V = \frac{4}{3} \pi a b c, \label{volume_abc}
\end{equation}
where $a$, $b$, and $c$ are the semi-axes of the ellipsoid. Depending on the droplet orientation, the axis of symmetry was assumed along either the major or minor axis. Specifically, for droplets oriented between $0^\circ-45^\circ$ and $135^\circ-180^\circ$, symmetry was considered about the minor axis (oblate shape), while for other angles, symmetry was taken about the major axis (prolate shape). Finally, the equivalent spherical diameter $(d)$ of the droplet was computed using:
\begin{equation} 
d = 2 \left( \frac{3V}{4\pi} \right)^{1/3}, \label{volume_4pi}
\end{equation}
which represents the diameter of a sphere having the same volume as the measured droplet. 

\section{Results and discussion} \label{sec:dis}

\begin{figure}
\centering
{\large (a)} \\
\includegraphics[width=0.75\textwidth]{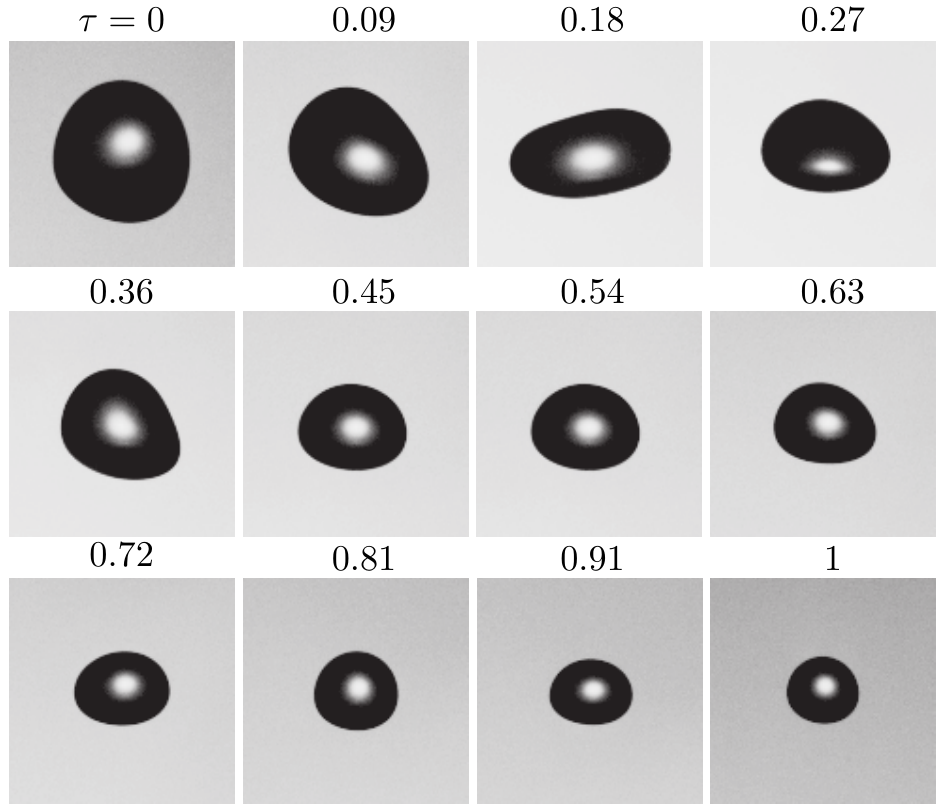} \\
\vspace{5mm}
{\large (b)} \\
\includegraphics[width=0.75\textwidth]{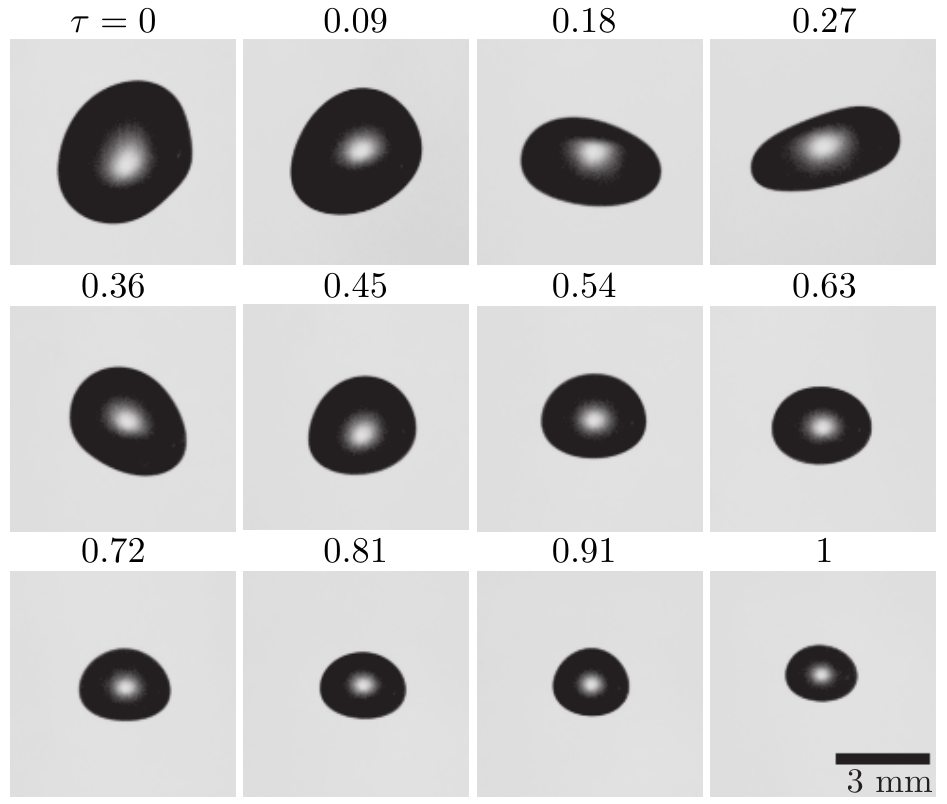}
\caption{Temporal evolution of the morphology of a levitated droplet undergoing evaporation in an upward airstream captured from two orthogonal views: (a) $yz$ - view and (b) $xy$ - view, using synchronized cameras as shown in figure \ref{schematic}. Here, $d_{0} = 4.0\pm 0.3$ mm, $RH = 10\%$ and $T = 30^{\circ}$C. The normalized time is given by $\tau = t/t_{80}$, where $t$ is the instantaneous time and $t_{80}$ denotes the time at which 80\% of the initial volume of the droplet has evaporated. The corresponding combined video showing both views is provided as a Supplementary Movie.}
\label{fig:shape}
\end{figure}

We examine the evaporation dynamics of water droplets of varying sizes under different temperature and humidity conditions in the Z-type vertical wind tunnel described in \S\ref{sec:expt}. The droplet is suspended in the test section (figure \ref{schematic}) by an upward-moving, conditioned airstream whose average velocity matches the terminal velocity of the droplet, so that the drag force balances its weight. Subsequently, based on the experimental findings, a theoretical model is developed to predict the evaporation time of a droplet under varying initial size, temperature, and humidity conditions. We begin the presentation of our results by demonstrating the temporal evolution of the evaporation dynamics for a typical droplet with an initial diameter of $d_0 = 4$ mm at $RH = 10\%$ and $T = 30^{\circ}$C, captured from two orthogonal views, as shown in figures~\ref{fig:shape}(a,b). The normalized time, $\tau = t/t_{80}$, is indicated at the top of each panel, where $t$ is the instantaneous time and $t_{80}$ denotes the time at which 80\% of the initial volume of the droplet has evaporated. In the present experiments, the dynamics are recorded up to $t = t_{80}$, beyond which further tracking becomes unreliable as the droplet gradually decreases in size over time due to evaporation. As seen in figures~\ref{fig:shape}(a) and \ref{fig:shape}(b), the droplet also undergoes complex morphological changes during evaporation, resulting from the interplay between inertial and surface tension forces. Inertia drives the shape oscillations between oblate and prolate forms, while surface tension acts to restore the droplet to a spherical shape. These oscillations significantly affect the evaporation mass flux and, consequently, the total evaporation time. By conducting a theoretical analysis, \cite{tonini2014evaporation} also reported that the evaporation rate of non-spherical droplets is strongly influenced by their shape oscillations. Thus, before analyzing the evaporation dynamics, we first examine the shape oscillations of the droplet under different temperatures, humidities, and initial sizes.

\subsection{Droplet oscillations}

\begin{figure}
\centering
\includegraphics[width=0.75\textwidth]{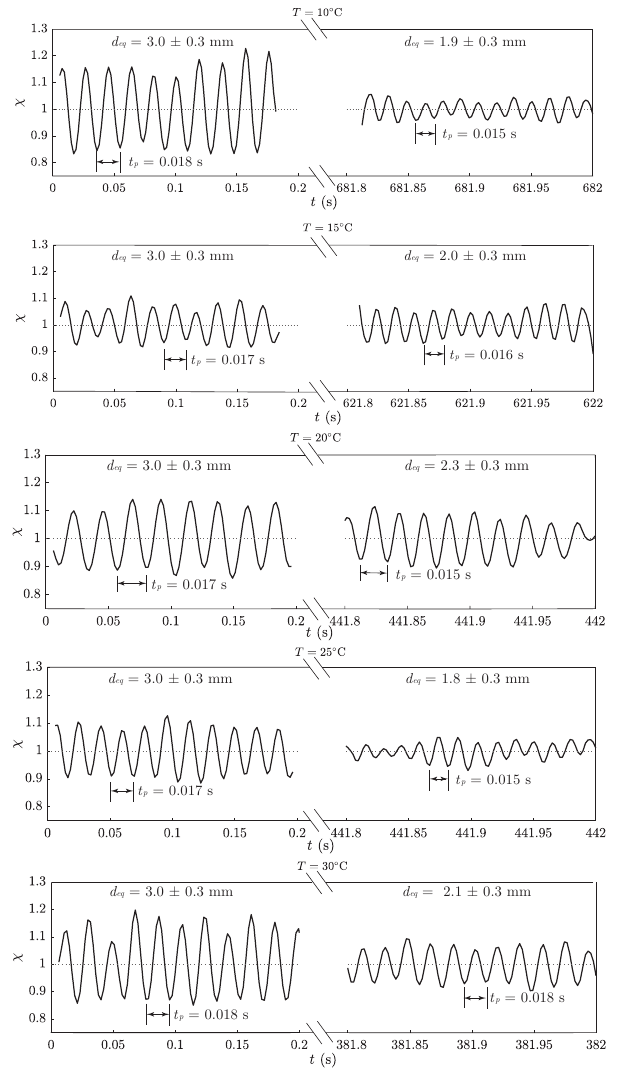}
\caption{Temporal evolution of the normalized axis ratio ($\chi$) of the oscillating droplet at different temperatures ($T$) during the early and later stages of evaporation. The remaining parameters are $d_0 = 3.0 \pm 0.3$ mm and $RH = 50\%$. The time period ($t_p$) and the equivalent droplet diameter ($d_{eq}$) corresponding to the early and later stages are indicated in the respective panels.}
\label{Shape_T}
\end{figure}

\begin{figure}
\centering
\includegraphics[width=0.76\textwidth]{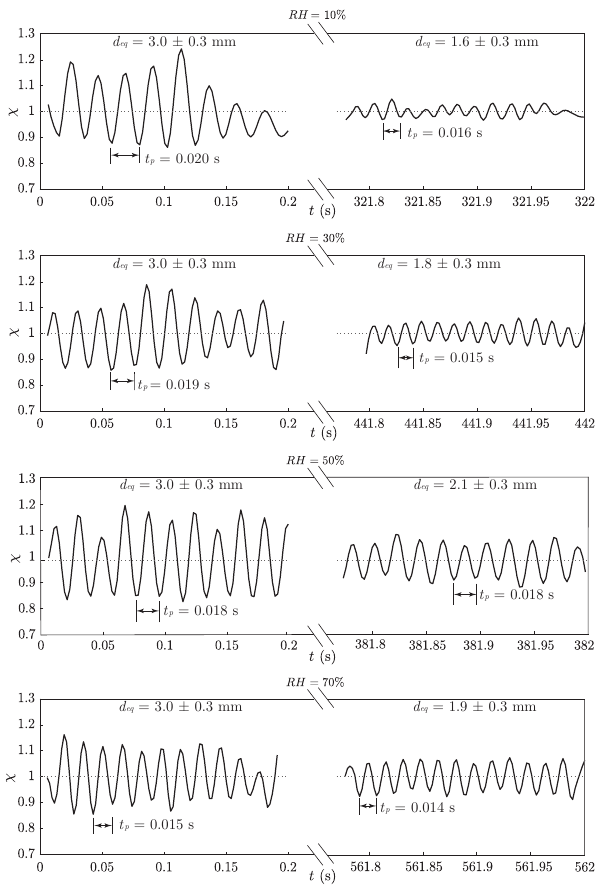}
\caption{Temporal evolution of the normalized axis ratio ($\chi$) of the oscillating droplet at different relative humidity ($RH$) levels during the early and later stages of evaporation. The remaining parameters are $d_0 = 3.0 \pm 0.3$ mm and $T = 30^\circ$C. The time period ($t_p$) and the equivalent droplet diameter ($d_{eq}$) corresponding to the early and later stages are indicated in the respective panels.}
\label{Shape_RH}
\end{figure}

\begin{figure}
\centering
\includegraphics[width=0.76\textwidth]{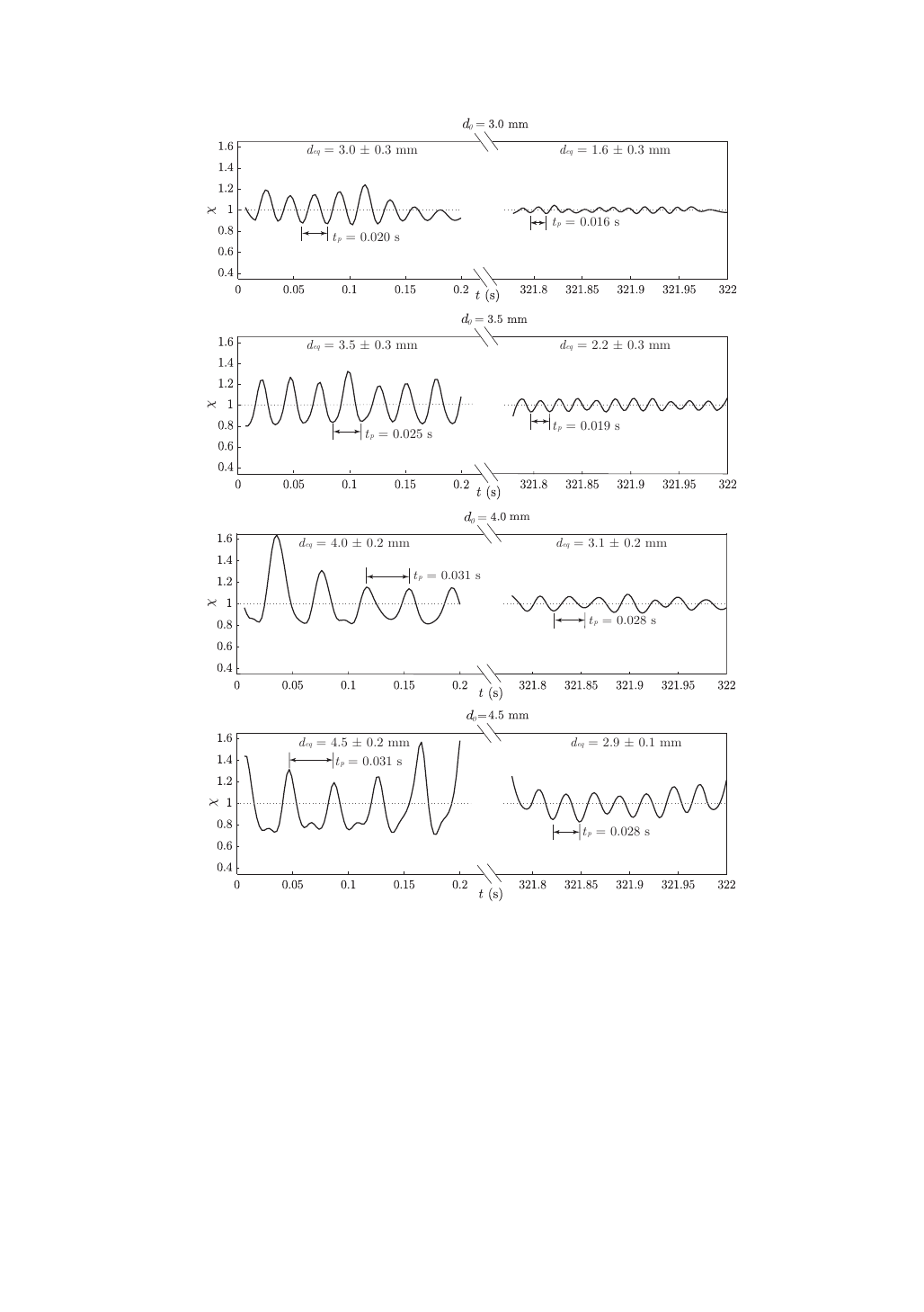}
\caption{Temporal evolution of the normalized axis ratio ($\chi$) of the oscillating droplet for different initial droplet diameters ($d_0$) during the early and later stages of evaporation. The remaining parameters are $RH = 10\%$ and $T = 30^\circ$C. The time period ($t_p$) and the equivalent droplet diameter ($d_{eq}$) corresponding to the early and later stages are indicated in the respective panels.}
\label{Shape_d}
\end{figure}

Figures~\ref{fig:shape}(a,b) illustrate that larger droplets ($d > 1$ mm) undergo complex deformations and exhibit persistent shape oscillations, similar to those observed in raindrops during their descent \citep{pruppacher1971semi,beard1984raindrop,chandrasekar1988axis,szakall2009wind}. Following the procedure discussed in \S\ref{sec:post-processing}, we compute the major and minor axes of the droplet undergoing periodic shape oscillations while evaporating \citep{agrawal2020experimental}. These oscillations are quantified using the normalised axis ratio ($\chi$), which is defined as
\begin{equation} \label{axis_ratio}
\chi = \left( \frac{a}{b} \right) \Bigg/ \int_0^1 \left( \frac{a}{b} \right) d\tau.
\end{equation}

Figures \ref{Shape_T}, \ref{Shape_RH}, and \ref{Shape_d} present the temporal evolution of the normalized axis ratio ($\chi$) of a levitated droplet, highlighting the effect of temperature, relative humidity, and initial droplet diameter, respectively, on shape oscillations during evaporation. Figure~\ref{Shape_T} depicts the temporal variation of the normalized axis ratio ($\chi$) of a levitated droplet at different temperatures ($T$) during the early and later stages of evaporation. In this experiment, the temperature in the test section is varied from $10^{\circ}$C to $30^{\circ}$C, while the relative humidity is held constant at $RH = 50\%$, and the initial droplet diameter is maintained at $d_0 = 3.0 \pm 0.3$ mm. It can be seen that the droplet exhibits periodic shape oscillations about $\chi = 1$, which corresponds to a spherical shape where the major and minor axes are equal ($a = b$). Figure \ref{Shape_T} reveals that larger deformations (i.e., higher values of $\chi$) occur during the early stages of evaporation due to greater inertia when the droplet is larger, while the deformations reduce (value of $\chi$ reduces significantly) in the later stages when the droplet becomes smaller due to evaporation and the surface tension becomes more dominant. Despite these changes in deformation amplitude, the time period of shape oscillations decreases slightly with time. For reference, the equivalent droplet diameter ($d_{eq}$) and the oscillation period ($t_p$) corresponding to the early and later stages are indicated at the top and bottom of each panel, respectively. Assuming $\rho_w \gg \rho_a$, the theoretical time period of shape oscillations for the fundamental mode ($n = 2$) simplifies to:
\begin{equation} \label{tp}
t_{p} = 2\pi \sqrt{\frac{\rho_wd_{0}^{3}}{64\sigma}}.
\end{equation}
It is noteworthy that the experimentally observed time period ($t_p$) of oscillations is consistent with the theoretical prediction given in eq.~\eqref{tp}. Figure~\ref{Shape_RH}  examines the effect of relative humidity, with $RH$ varied while maintaining the temperature at $T = 30^\circ$C and the droplet diameter at $d_0 = 3.0 \pm 0.3$ mm. A similar trend is observed, with pronounced shape oscillations during the early stages, while the overall amplitude of oscillations diminishes in the later stages of evaporation as the droplet shrinks. It can be seen that as the humidity level decreases, droplet evaporation accelerates, leading to a reduction in the oscillation time period over time, with this effect being more pronounced under low humidity conditions. In figure \ref{Shape_d}, the impact of initial droplet diameter is studied by varying $d_0$ from 3.0 to 4.5 mm at fixed $RH = 10\%$ and $T = 30^\circ$C. As expected, larger droplets exhibit more pronounced oscillations in the early stages due to the increased influence of inertial forces and reduced surface tension effects. As evaporation progresses, these deformations decrease. Close inspection of figure \ref{Shape_d} also reveals that the oscillation time period is found to increase with droplet size, which is in good agreement with theoretical predictions. 

\begin{figure}
\centering
\hspace{1.0 cm} {\large (a)} \hspace{6.2 cm} {\large (b)} \\
\includegraphics[width=0.488\textwidth]{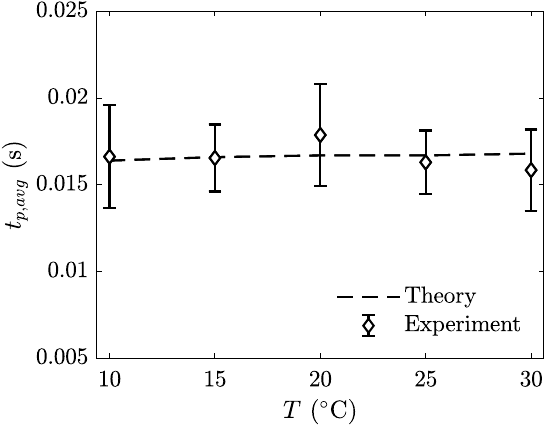} \hspace{0.5mm}
\includegraphics[width=0.488\textwidth]{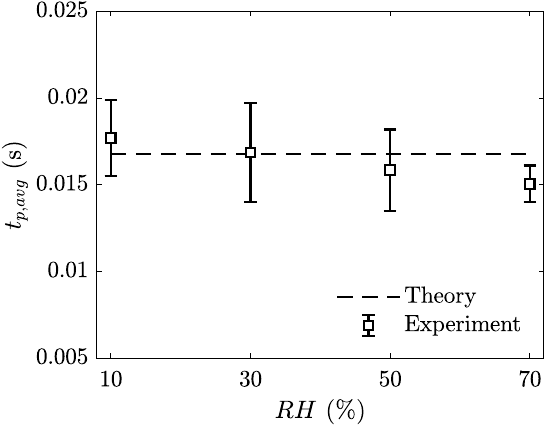} \\
\hspace{1.0 cm} {\large (c)}\\
\includegraphics[width=0.488\textwidth]{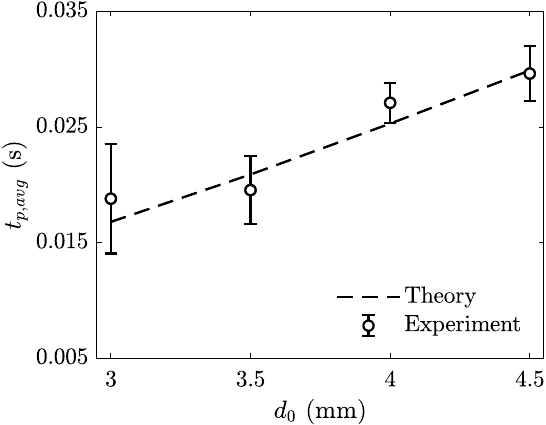}
\caption{Comparison of the time period of droplet oscillations, averaged over its lifetime ($t_{p,avg} = \int_0^1 t_p d \tau$), obtained experimentally, with the theoretical prediction using eq.~\eqref{tp}, for (a) varying temperature ($T$) at $RH = 50\%$ and $d_0 = 3.0 \pm 0.3$ mm, (b) varying relative humidity ($RH$) at $T = 30^\circ$C and $d_0 = 3.0 \pm 0.3$ mm, and (c) varying initial droplet diameter ($d_0$) at $RH = 10\%$ and $T = 30^\circ$C. The error bars indicate the standard deviation of the oscillation time period measured over the lifetime of the droplet.}
\label{time_period}
\end{figure}

Figure~\ref{time_period} compares the theoretically predicted oscillation time period ($t_p$), derived from eq. \eqref{tp}, with the experimentally measured average oscillation time period ($t_{p,avg} = \int_0^1 t_p d \tau$) under different conditions. Figure~\ref{time_period}(a) presents the variation of $t_p$ with temperature ($T$), keeping relative humidity fixed at $RH = 50\%$ and the initial droplet diameter at $d_0 = 3.0 \pm 0.3$ mm. Figure~\ref{time_period}(b) examines the effect of relative humidity ($RH$) at a constant temperature of $T = 30^\circ$C and the same droplet diameter, while figure~\ref{time_period}(c) shows the influence of initial droplet diameter ($d_0$) at fixed $RH = 10\%$ and $T = 30^\circ$C. In all cases, the experimental time period of oscillations is averaged over the lifetime of the droplet, with error bars indicating the standard deviation measured over the same duration. We observe that $t_{p,\mathrm{avg}}$ changes only slightly with temperature and humidity for the conditions considered in figure~\ref{time_period}(a,b). However, figure~\ref{time_period}(c) shows that $t_{p,\mathrm{avg}}$ increases with increasing $d_0$, indicating a strong dependence on droplet diameter, which is in good agreement with theoretical predictions. This theoretical timescale is further employed in the analysis of evaporation dynamics in the subsequent section. Next, we examine the evaporation dynamics by analyzing the temporal variation of the volume and diameter of the levitated droplet under different conditions.

\subsection{Evaporation dynamics}

\begin{figure}
\centering
\hspace{0.7cm} {\large (a)} \hspace{6.2 cm} {\large (b)} \\
\includegraphics[width=0.48\textwidth]{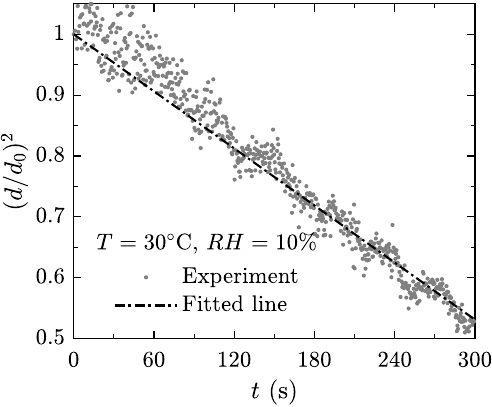} \hspace{1mm}
\includegraphics[width=0.48\textwidth]{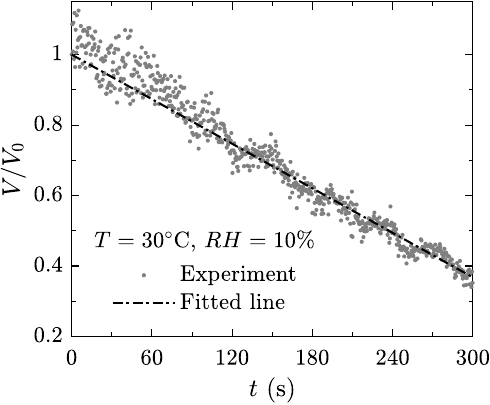}
\caption{Temporal evolution of (a) $(d/d_0)^2$ and (b) $V/V_0$ of an evaporating droplet for a representative set of parameters ($d_0 = 3.0 \pm 0.3$ mm, $RH = 10\%$, and $T = 30^\circ$C), obtained from data captured from high-speed imaging system. The dash-dotted line represents the best fit to the experimental data points.}
\label{fig:dia_vol_1}
\end{figure}

Figures \ref{fig:dia_vol_1}(a,b) illustrate the procedure employed to obtain the temporal evolution of the normalized droplet diameter $(d/d_0)^2$ and volume ($V/V_0$) from high-speed imaging data for a representative case ($RH = 10\%$, $T = 30^\circ$C, $d_{0} = 3.0 \pm 0.3$ mm). In figures~\ref{fig:dia_vol_1}(a) and \ref{fig:dia_vol_1}(b), the dots represent experimental estimates of $(d/d_0)^2$ and $V/V_0$, respectively, obtained from synchronized orthogonal camera views at each time instant, as calculated using the procedure described in \S\ref{sec:post-processing}. The solid curves indicate the corresponding best-fit trends. The nearly linear decrease in both squared of the normalized diameter and normalized volume with time indicates a steady evaporation rate under these conditions. These representative plots serve as a reference for analyzing the influence of different parameters in subsequent figures. For all other cases presented in the following, only the fitted curves are shown to depict the temporal variation of droplet diameter and volume.

\begin{figure}
\centering
\hspace{0.6 cm} {\large (a)} \hspace{6.2 cm} {\large (b)} \\
\includegraphics[width=0.485\textwidth]{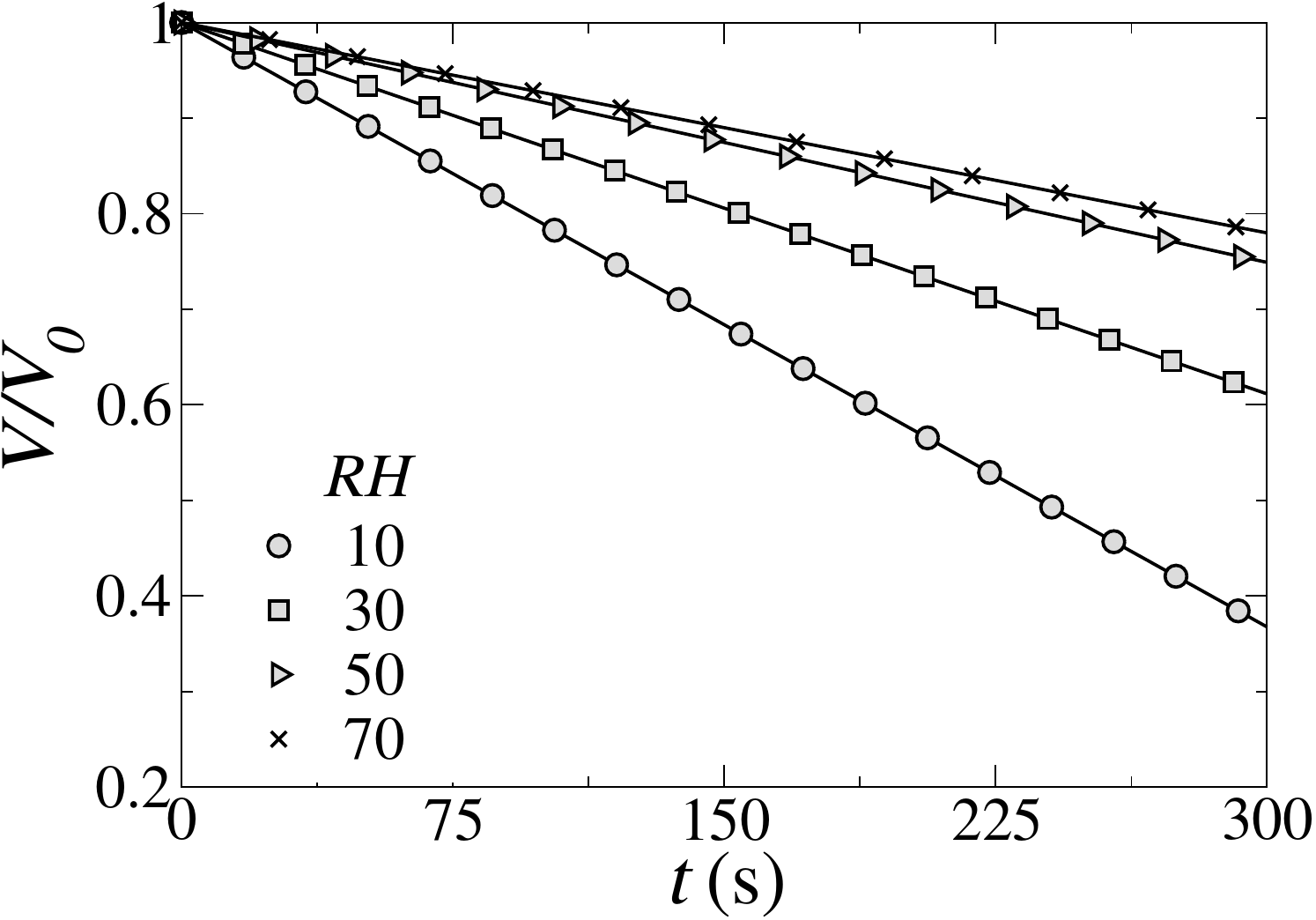} \hspace{0.5mm}
\includegraphics[width=0.485\textwidth]{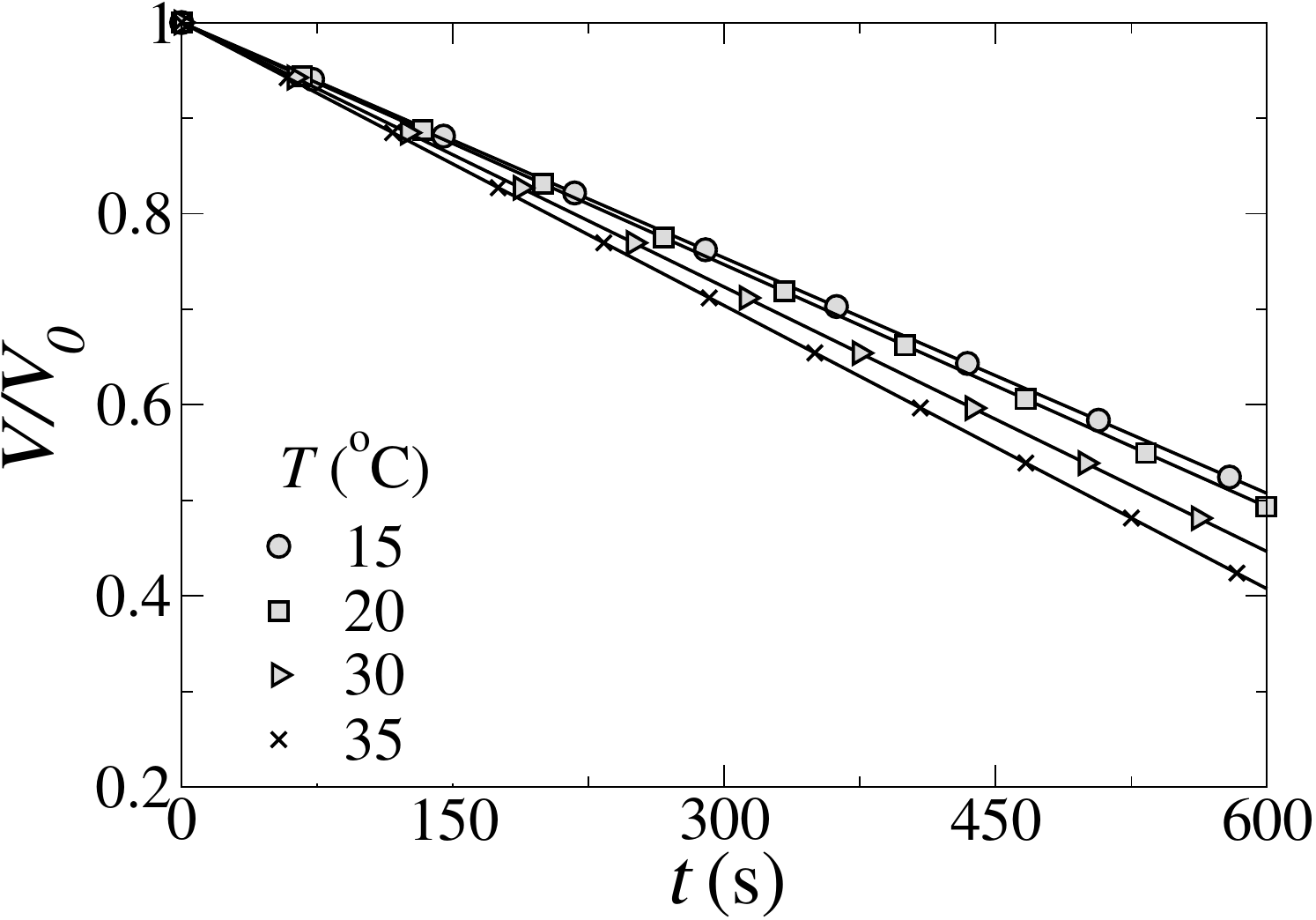} \\
\hspace{0.6 cm} {\large (c)} \\
\includegraphics[width=0.485\textwidth]{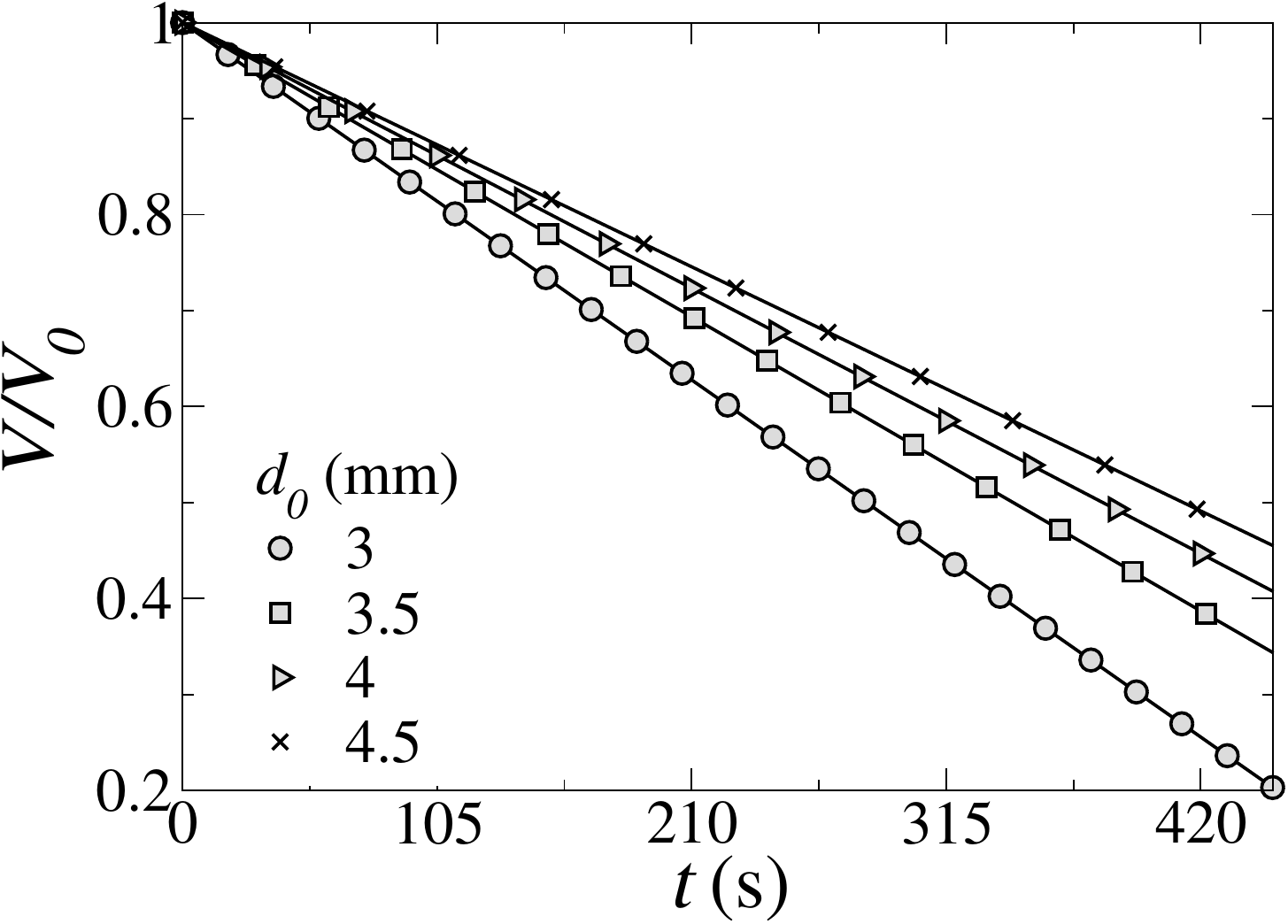}
\caption{Temporal evolution of the normalized droplet volume ($V/V_0$) under various experimental conditions: (a) effect of relative humidity ($RH$) at $T = 30^\circ$C and $d_0 = 3.0 \pm 0.3$ mm, (b) effect of temperature ($T$) at $RH = 50\%$ and $d_0 = 3.0 \pm 0.3$ mm, and (c) effect of initial droplet diameter ($d_0$) at $T = 30^\circ$C and $RH = 10\%$.} 
\label{fig:dia_vol_3}
\end{figure}

Next, we perform a systematic parametric study to investigate the effects of temperature ($T$), relative humidity ($RH$), and initial droplet diameter ($d_0$) on the evaporation behaviour by analyzing the temporal evolution of the normalized droplet volume ($V/V_0$) for different sets of parameters. Figure~\ref{fig:dia_vol_3}(a) illustrates the effect of relative humidity ($RH$) on the temporal evolution of the normalized droplet volume ($V/V_{0}$) for $T = 30^\circ$C and $d_0 = 3.0 \pm 0.3$ mm. Each curve represents a best-fit line to the experimental data for different humidity levels. It can be seen that increasing the value of $RH$ slows down the evaporation process. This trend is physically consistent, as increasing $RH$ elevates the ambient water vapour mass fraction, thereby reducing the concentration gradient between the droplet surface and the surrounding air. This diminished gradient weakens the mass transfer driving force, leading to slower evaporation. The opposite occurs at lower $RH$, where the stronger gradient promotes faster evaporation. Figure~\ref{fig:dia_vol_3}(b) shows the influence of temperature ($T$) on the evaporation rate, keeping $RH = 50\%$ and $d_{0} = 3.0 \pm 0.3$ mm fixed. As expected, higher temperatures accelerate evaporation, while lower temperatures result in slower rates. This behaviour is attributed to the increase in saturation vapour pressure at the droplet surface with temperature, which enhances the vapour concentration gradient and facilitates greater mass transfer. In contrast, at lower temperatures, the reduced saturation pressure weakens this gradient, thereby suppressing evaporation. Figure~\ref{fig:dia_vol_3}(c) presents the effect of varying initial droplet diameter ($d_0$) at constant $T = 30^\circ$C and $RH = 10\%$. The results reveal that smaller droplets evaporate more rapidly than larger ones. This is primarily due to the higher surface-area-to-volume ratio of smaller droplets, which exposes a greater fraction of their mass to the ambient environment, enhancing the mass loss rate. Although the vapour concentration gradient is similar across different droplet sizes under fixed ambient conditions, the total surface area available for diffusion increases less rapidly than volume with increasing size. Consequently, larger droplets evaporate more slowly, retaining their mass over a longer duration. These observations confirm the strong dependence of evaporation dynamics on $RH$, $T$, and $d_0$. In the following section, we analyze the temporal variation of $d/d_0$, develop a theoretical model to gain deeper insight into the underlying physical mechanisms, and compare its predictions with the experimental results.

\subsection{Theoretical modelling}

A brief discussion of the classical theoretical model based on the $d^2$-law is provided in Appendix~\ref{appendix}. This model describes the evaporation rate of a stationary spherical droplet in a quiescent environment, assuming a constant evaporation flux. However, this framework fails to accurately capture the evaporation dynamics of a levitated droplet exposed to an airstream, where shape oscillations and convective effects play a dominant role. As shown in figures~\ref{fig:dia_vol_4}(a–c), the theoretical predictions based on the classical $d^2$-law deviate considerably from the experimental observations for a levitated droplet under different conditions. Therefore, extending the classical model to incorporate these additional effects is essential for a more realistic and accurate description of the evaporation process under varying environmental conditions. To address the discrepancies arising from convective transport and droplet shape oscillations, a modified evaporation model is presented in the following section.

\subsubsection{Modified evaporation model} \label{modified_model}
 
\begin{figure}
\centering
\includegraphics[width=0.60\textwidth]{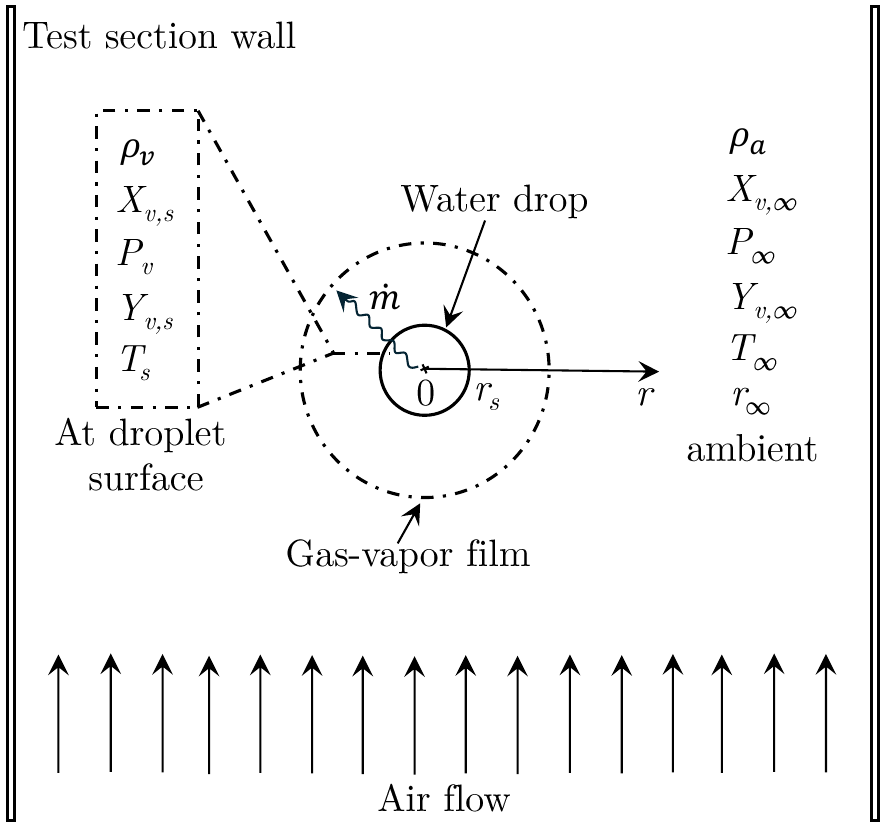}
\caption{Schematic representation of the evaporation mechanism for a droplet surrounded by a gas-vapour film in a gaseous medium subjected to forced convection. Here, $\dot{m}$ denotes the mass flow rate from the droplet surface into the surrounding gas-vapour region along the radial direction ($r$). The properties and parameters at the surface of the droplet ($r = r_s$) and at the ambient ($r\rightarrow r_{\infty}$) have been illustrated.}
\label{drop}
\end{figure}

The evaporation rate of water droplets in quiescent air is primarily governed by the transfer of water vapour and heat between the droplet surface and the surrounding environment. However, in the presence of relative motion between the air and the droplet, as observed in cloud droplets or raindrops falling under gravity, the airflow around the droplet significantly enhances vapour and heat transfer. Moreover, under realistic atmospheric conditions, these droplets exhibit shape oscillations and experience variations in ambient temperature and humidity. To account for the effects of convection, shape oscillations, and environmental variability, we propose a modified evaporation model that incorporates the combined influence of ambient air motion, temperature, and relative humidity on the evaporation dynamics. Note that while a few studies have proposed modifications to the classical evaporation model under simplified conditions \citep{ranz1952evaporation,pinheiro2019evaluation,nugraha2022sherwood,beji2018detailed,tonini2014evaporation}, we build upon and enhance these models to develop a more comprehensive and physically consistent framework suitable for raindrop-like conditions. In Appendix \ref{appendix_PIV}, Particle Image Velocimetry (PIV) measurements are provided to quantify the velocity field within the wind-tunnel test section. As shown in figure \ref{vel_prof}(a–c), the velocity distribution remains nearly uniform away from the centerline, although a small dip (velocity well) is observed near the central region. Across all flow rates examined, this velocity well is approximately 30\% lower than the average velocity and spans only about 20\% of the test-section width. Due to its limited spatial extent, we assume that the velocity well has a minimal effect on the evaporation dynamics of the suspended droplet, and adopt a uniform velocity profile for theoretical modelling. Figure~\ref{drop} presents a schematic of the evaporation mechanism within the wind tunnel test section (shown in figure~\ref{schematic}), where a droplet is levitated by an upward airstream such that the drag force balances its weight. The solid circular boundary denotes the droplet surface, while the surrounding dashed region represents the gas–vapour film. The region beyond this film corresponds to the ambient environment. Physical properties vary along the radial direction $r$, and the rate of evaporation, denoted by $\dot{m}$, is illustrated by a wavy arrow directed outward from the droplet surface into the gas–vapour region. The modeling is based on the following assumptions: (i) the evaporation process is quasi-steady, (ii) the droplet temperature remains spatially uniform, (iii) the vapor mass fraction at the droplet surface is equal to its saturation value, (iv) physical properties such as saturated vapor density and binary diffusivity are constant, and (v) the droplet surface temperature is equal to the ambient air temperature. \\
 
\noindent {\it (i) Convection effects:} 
The evaporation rate of a droplet accounting for convective effects can be expressed as \citep{pinheiro2019evaluation}:
\begin{equation} \label{mass_flux}
\dot{m} = \pi d {D_{v} \rho_v} \Sh \ln(1 + B),
\end{equation}
where $D_v$ is the diffusion coefficient of water vapor in air, $\rho_v$ is the saturated vapor density, and $B$ is the Spalding mass transfer number, which is defined as \citep{spalding1960standard}:
\begin{equation} \label{mass_tran_number}
B = \frac{Y_{v,s} - Y_{v,\infty}}{1 - Y_{v,s}}.
\end{equation}
Here, $Y_{v,s}$ and $Y_{v,\infty}$ denote the mass fractions of water vapor at the droplet surface (saturation) and in the ambient environment, respectively. The expressions for $Y_{v,s}$ and $Y_{v,\infty}$ are provided in eqs.~\eqref{yvs} and \eqref{yvinfi}, respectively. Additional details are presented in Appendix~\ref{appendix}.

It is to be noted that, in addition to the Spalding mass transfer number ($B$), the Spalding heat transfer number ($B_T$) is also an important parameter in droplet heating and combustion modeling, particularly when the surface temperature ($T_s$) differs significantly from the ambient temperature ($T_{\infty}$) \citep{sazhin2006advanced, sazhin2014two}. In the present study, we assume $T_s = T_{wb}$, where $T_{wb}$ is the wet-bulb temperature under steady-state equilibrium condition. Thus, $B_T$ is defined as
\begin{equation} \label{Spalding_heat}
B_T = \frac{C_{pv}(T_{\infty} - T_{wb})}{L(T_{wb}) - (\dot{q_d}/\dot{m_d})},
\end{equation}
where $C_{pv}$ is the specific heat of the vapor, $L(T_{wb})$ is the latent heat of vaporization, and $\dot{q_d}$ and $\dot{m_d}$ denote the heat and mass transfer rates across the liquid–vapor interface, respectively. However, unlike combustion or spray-heating scenarios, our experiments involve pure water droplets evaporating in a controlled aerodynamic levitation setup, where the ambient temperature ($T_{\infty}$) remains relatively low (below $50^\circ$C) and no internal heat source is present. The Spalding heat transfer number $(B_T)$ has been evaluated using a psychrometric chart at the specified ambient temperature and relative humidity. Our calculations show that $B_T \ll B$ over the range of experimental conditions considered. The dominant contribution to evaporation arises from vapor concentration-driven mass transfer enhanced by forced convection, which is already captured in the Sherwood-number formulation used in our model. Consequently, we only employ the Spalding mass transfer number ($B$) to describe the evaporation process, as it inherently captures the coupled influence of ambient temperature and relative humidity through vapor concentration gradients at the droplet interface. This formulation is well established for isothermal or near-isothermal droplet evaporation and provides an accurate description of the physical process without introducing additional parameters \citep{yarin1999evaporation, maruyama2020evaporation, mitsuno2025evaporation}.

In eq. \eqref{mass_flux}, the Sherwood number ($\Sh$) is introduced to capture the enhancement in mass transfer due to airflow around the droplet. It characterizes the convective mass transfer of vapour from the droplet surface to the surrounding air. The Sherwood number is defined as:
\begin{equation} \label{Sherwood_def}
\Sh = \frac{h_{m} d}{{D_{v}}},
\end{equation}
where $h_m$ is the mass transfer coefficient. To evaluate the Sherwood number ($\Sh$), several correlations have been proposed in the literature \citep{nugraha2022sherwood,beji2018detailed}, with the correlation of \cite{ranz1952evaporation} being among the most widely used. However, these traditional formulations typically express $\Sh$ as a function of the Reynolds number ($Re$) and the Schmidt number ($\Sc$), while neglecting the effects of ambient temperature and relative humidity. In the present study, we propose a modified correlation for the Sherwood number that not only retains its dependence on $Re$ and $\Sc$, but also explicitly incorporates the influence of ambient temperature ($T$) and relative humidity ($RH$) on the evaporation process.

We aim to model the Sherwood number ($\Sh$) based on various parameters obtained from experiments. For each measurement (denoted by $i$), the available data include the Reynolds number ($Re_i$), Schmidt number ($\Sc_i$), relative humidity ($RH_i$), a temperature ratio ${\left( \frac{T_{\infty }-T_{\text{ref}}}{T_{\text{ref}}} \right)}_i$, and the experimentally measured Sherwood number ($\Sh_{\text{actual},i}$). Here, $T_{\infty}$ and $T_{\text{ref}} = 273,\mathrm{K}$ represent the ambient air temperature and the reference temperature, respectively. In the present study, the experimental Sherwood number is evaluated as \citep{abramzon1989droplet}:
\begin{equation} \label{Sh_actual}
\Sh_{\text{actual}} = \frac{m^{\prime\prime}_{\text{actual}} d_0}{{\rho_v D_v} ({Y_{v,s}} - {Y_{v,\infty}})},
\end{equation}
where $m^{\prime\prime}_{\text{actual}}$ is the experimental mass flux, which is evaluated as:
\begin{equation} \label{evap_actual}
m^{\prime\prime}_{\text{actual}} = -{\rho_v} \frac{\mathrm{d} r_s}{\mathrm{d} t}.
\end{equation}
where $(\mathrm{d}r_s/\mathrm{d}t)$ represents the rate of change of the droplet radius. The Reynolds and Schmidt numbers are determined using
\begin{equation} \label{ReSc_num}
Re = \frac{\rho_a V_t d_0}{\mu_a}, \hspace{2mm} \text{and} \hspace{2mm}
\Sc = \frac{\mu_a}{\rho_a {D_v}},
\end{equation}
where $\rho_a$ and $\mu_a$ are the density and dynamic viscosity of the ambient air, respectively, and $V_t$ is the terminal velocity of the droplet.

The proposed model assumes that the Sherwood number follows a power-law relationship given by:
\begin{equation} \label{sh_predicted}
\Sh_{\text{predicted},i} = 2 + A ~ Re_i^{p}  ~ \Sc_i^{q} ~RH_i^{r} \left( \frac{T_{\infty} - T_{\text{ref}}}{T_{\text{ref}}} \right)_i^{{s}},
\end{equation}
where $\Sh_{\text{predicted}}$ is the predicted Sherwood number, $A$ is a scaling coefficient, and $p$, $q$, $r$, and $s$ are the exponents corresponding to the Reynolds number, Schmidt number, relative humidity, and temperature ratio, respectively. The unknowns to be determined through optimization are $A$, $p$, $q$, $r$, and $s$. The objective is to minimize the mean absolute percentage error (MAPE) between the predicted and actual Sherwood numbers, defined as:
\begin{equation} \label{mape_eqn}
\text{MAPE} = \frac{100}{n} \sum_{i=1}^{n} \left| \frac{\Sh_{\text{actual},i} - \Sh_{\text{predicted},i}}{\Sh_{\text{actual},i}} \right|,
\end{equation}
where $n$ is the total number of observations. The optimization problem is thus formulated as:
\begin{equation} \label{min_eqn}
\min_{A, p, q, r, s} \quad \text{MAPE}(A, p, q, r, s).
\end{equation}

To solve this optimization problem, an initial guess for the parameters $(A, p, q, r, s)$ is chosen, and a derivative-free optimization algorithm (Nelder–Mead simplex method described in \cite{lagarias1998convergence}) is employed to iteratively adjust the parameters. After convergence, the optimized values $(A_{\text{opt}}, p_{\text{opt}}, q_{\text{opt}}, r_{\text{opt}}, s_{\text{opt}})$ are obtained. Using these optimized parameters, the predicted Sherwood numbers are recalculated for each observation as:
\begin{equation} \label{sh_pred_opt}
\Sh_{\text{predicted},i} = 2 + A_{\text{opt}} ~ Re_i^{p_{\text{opt}}} ~ \Sc_i^{q_{\text{opt}}} ~ RH_i^{r_{\text{opt}}} \left( \frac{T_{\infty} - T_{\text{ref}}}{T_{\text{ref}}} \right)_i^{{s_{\text{opt}}}}.
\end{equation}

Subsequently, the absolute error and percentage error for each observation are computed as:
\begin{equation} \label{abs_error}
\text{absolute error}_i = |\Sh_{\text{actual},i} - \Sh_{\text{predicted},i}|,
\end{equation}
\begin{equation} \label{perc_error}
\text{percentage error}_i = \left( \frac{\text{absolute error}_i}{\Sh_{\text{actual},i}} \right) \times 100.
\end{equation}
The final MAPE is then obtained by averaging the percentage errors across all observations. In the present study, the MAPE between $\Sh_{\text{actual}}$ and $\Sh_{\text{predicted}}$ is found to be $10.55\%$. The final correlation for $\Sh$ using the optimized power-law model is:
\begin{equation} \label{final_Sh}
\Sh = 2 + 0.04 Re^{0.50} \Sc^{0.33} RH^{-0.189} \left( \frac{T - T_{\text{ref}}}{T_{\text{ref}}} \right)^{-2.225}.
\end{equation}
Here, $\Sh = 2$ represents pure diffusion-driven evaporation in a quiescent medium, i.e., without convection. The additional term in eq.~\eqref{final_Sh} accounts for the enhancement in mass transfer due to convection around the droplet, which becomes significant for an evaporating droplet in moving airstreams.

The Sherwood number correlation obtained using the power-law model captures the key physical factors influencing droplet evaporation dynamics. However, it does not account for the effects of droplet shape oscillations that arise during evaporation. To address this limitation, we introduce an additional parameter, referred to as the shape factor, in the evaporation rate equation, which incorporates the influence of droplet oscillations. This modification is discussed in detail below. \\

\noindent {\it (ii) Effect of shape oscillations}: In order to account for the non-spherical shape of the droplet and its shape oscillations arising from the competition between inertia and surface tension, the evaporation rate $\dot{m}$ is expressed as \citep{tonini2014evaporation}:
\begin{equation} \label{mass_speroid}
\dot{m} = \pi d {D_{v} \rho_v} \Sh \ln(1 + B) f_{\text{shape}},
\end{equation}
where $f_{\text{shape}}$ is the ratio between the surface area of a spheroid ($A_{\text{spheroid}}$) and that of an iso-volumic spherical droplet ($A_{\text{sphere}}$). This ratio, which varies with time $t$, is approximated as \citep{tonini2014evaporation}:
\begin{equation} \label{area_ratio}
f_{\text{shape}} = \frac{A_{\text{spheroid}}}{A_{\text{sphere}}} = 1 + \Delta \beta \sin^2(\omega t),
\end{equation}
where $\Delta \beta$ is the dimensionless maximum excess surface area relative to the spherical state, and $\omega$ is the frequency of oscillation of the droplet (in rad/s), which can be evaluated as \citep{balla2019shape,lyubimov2011small,tonini2014evaporation}:
\begin{equation} \label{omega_eqn}
\omega = \sqrt{\frac{64 \sigma}{\rho_w d_0^3}}.
\end{equation}
This natural oscillation frequency is consistent with the experimentally observed frequency, calculated as $\omega = 2\pi/t_p$, as shown in figures \ref{Shape_T}–\ref{Shape_d}. From eq.~\eqref{area_ratio}, it is evident that the shape factor varies as the droplet oscillates. Therefore, it is important to consider the time-averaged shape factor over one oscillation period $t_p$. The average shape factor, $f_{\text{shape,avg}}$, is evaluated as:
\begin{equation} \label{fshape_int}
f_{\text{shape,avg}} = \frac{1}{t_p} \int_{0}^{t_p} f_{\text{shape}}  \mathrm{d}t = \frac{1}{t_p} \int_{0}^{t_p} \left(1 + \Delta \beta \sin^2(\omega t) \right) \mathrm{d}t,
\end{equation}
\begin{equation} \label{fshape_avg}
f_{\text{shape,avg}} = 1 + \frac{\Delta \beta}{2} \left(1 - \frac{\sin(2\omega t_p)}{2\omega t_p} \right).
\end{equation}
Substituting $t_p = 2\pi / \omega$, it further simplifies to:
\begin{equation} \label{fshape_avg_revised}
f_{\text{shape,avg}} = 1 + \frac{\Delta \beta}{2}.
\end{equation}
This can be used in eq.~\eqref{mass_speroid} to express the evaporation rate as:
\begin{equation} \label{mass_speroid_favg}
\dot{m} = \pi {d D_v \rho_v} \Sh \ln(1 + B) f_{\text{shape,avg}}.
\end{equation}
Solving eq.~\eqref{mass_speroid_favg} yields the modified evaporation rate constant ($K_{\text{mod}}$) as:
\begin{equation} \label{k}
K_{\text{mod}} = \frac{2 \rho_v D_v \Sh f_{\text{shape,avg}}}{\rho_w} \ln(1 + B),
\end{equation}
which leads to the modified $d^2$-law, given by
\begin{equation} \label{mod_dsquare}
d^2(t) = d_0^2 - K_{\text{mod}} t.
\end{equation}
The total evaporation time ($t_f$) of the droplet can be obtained by integrating eq.~\eqref{mass_speroid_favg} from $t = 0$ to $t = t_f$, which yields:
\begin{equation} \label{total_eva}
t_f = \frac{\rho_w d_0^2}{4 {D_v \rho_v} \Sh \ln(1 + B) f_{\text{shape,avg}}}.
\end{equation}
As discussed in \S\ref{sec:expt}, accurately capturing droplet dynamics becomes increasingly challenging as the droplet size diminishes. Therefore, in our experiments, we track the evaporation process only until 80\% of the initial volume of the droplet has evaporated, i.e., until $V = 0.8V_0$. This stage is denoted as $V_{80}$, and the corresponding droplet diameter is represented by $d_{80}$ ($\approx$ $0.58d_0$). The associated evaporation time, $t_{80}$, can be estimated by integrating eq.~\eqref{mass_speroid_favg} from $t = 0$ to $t = t_{80}$, and $d = d_0$ to $d = d_{80}$, resulting in:
\begin{equation} \label{t_80_int}
 \int_{d_0}^{d_{80}}(d)\mathrm{d}d = \frac{2 D_v \rho_{v} Sh f_{shape,avg}}{\rho_{w}} \ln(1+B)\int_{0}^{t_{80}}\mathrm{d}t.
\end{equation}
This further simplifies to
\begin{equation} \label{t80_eva}
t_{80} = \frac{0.7 \rho_w d_0^2}{4 {D_v \rho_v} \Sh \ln(1 + B) f_{\text{shape,avg}}}.
\end{equation}
Next, we compare the theoretical predictions obtained using the modified evaporation model, which accounts for convective effects and the temporal evolution of morphological deformation of the droplet, with the experimental results by examining the temporal variation of the droplet diameter and total evaporation time.

\subsubsection{Comparison with experimental results}

\begin{figure}
\centering
\hspace{0.6 cm} {\large (a)} \hspace{6.2 cm} {\large (b)} \\
\includegraphics[width=0.485\textwidth]{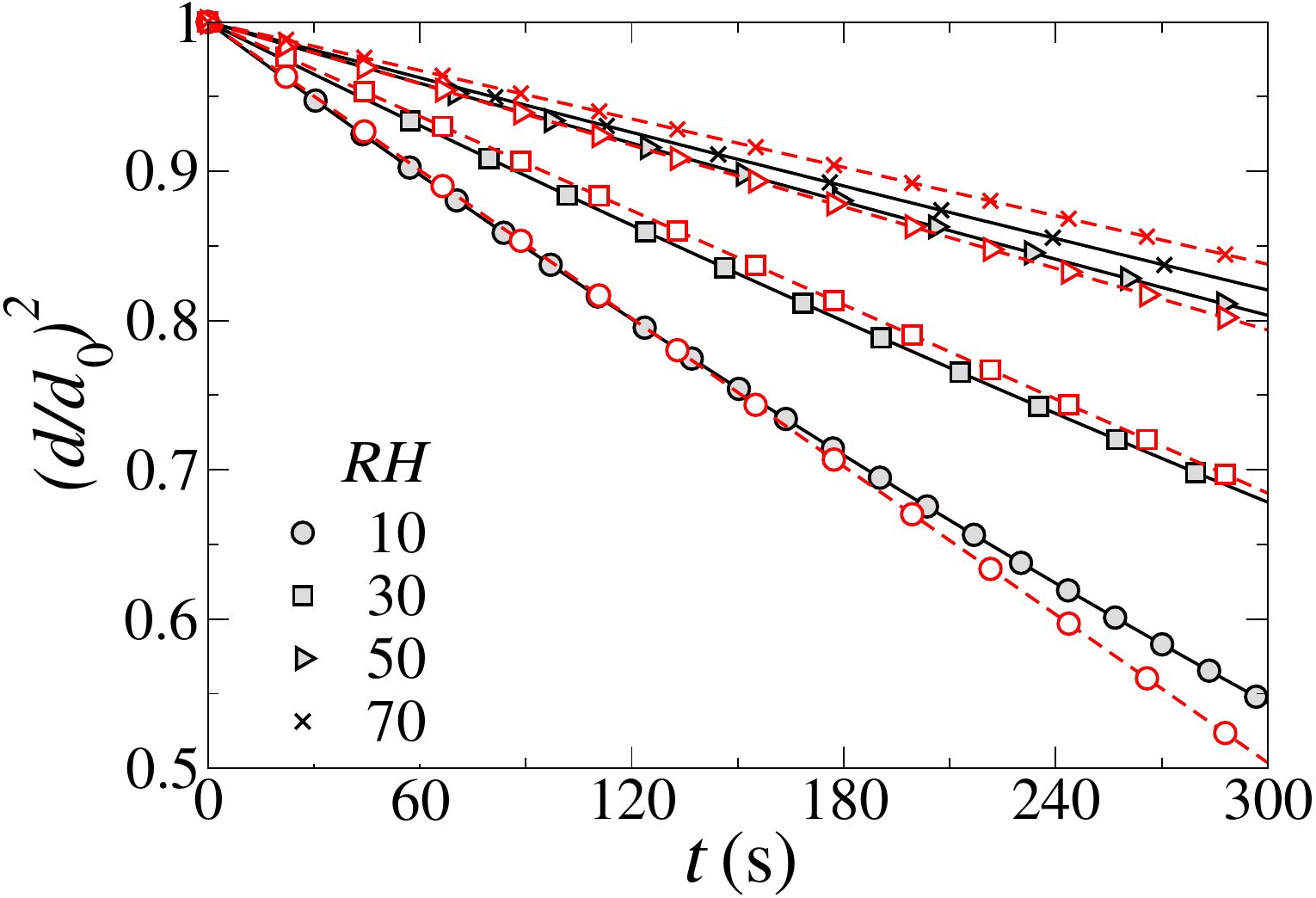} \hspace{0.5mm}
\includegraphics[width=0.485\textwidth]{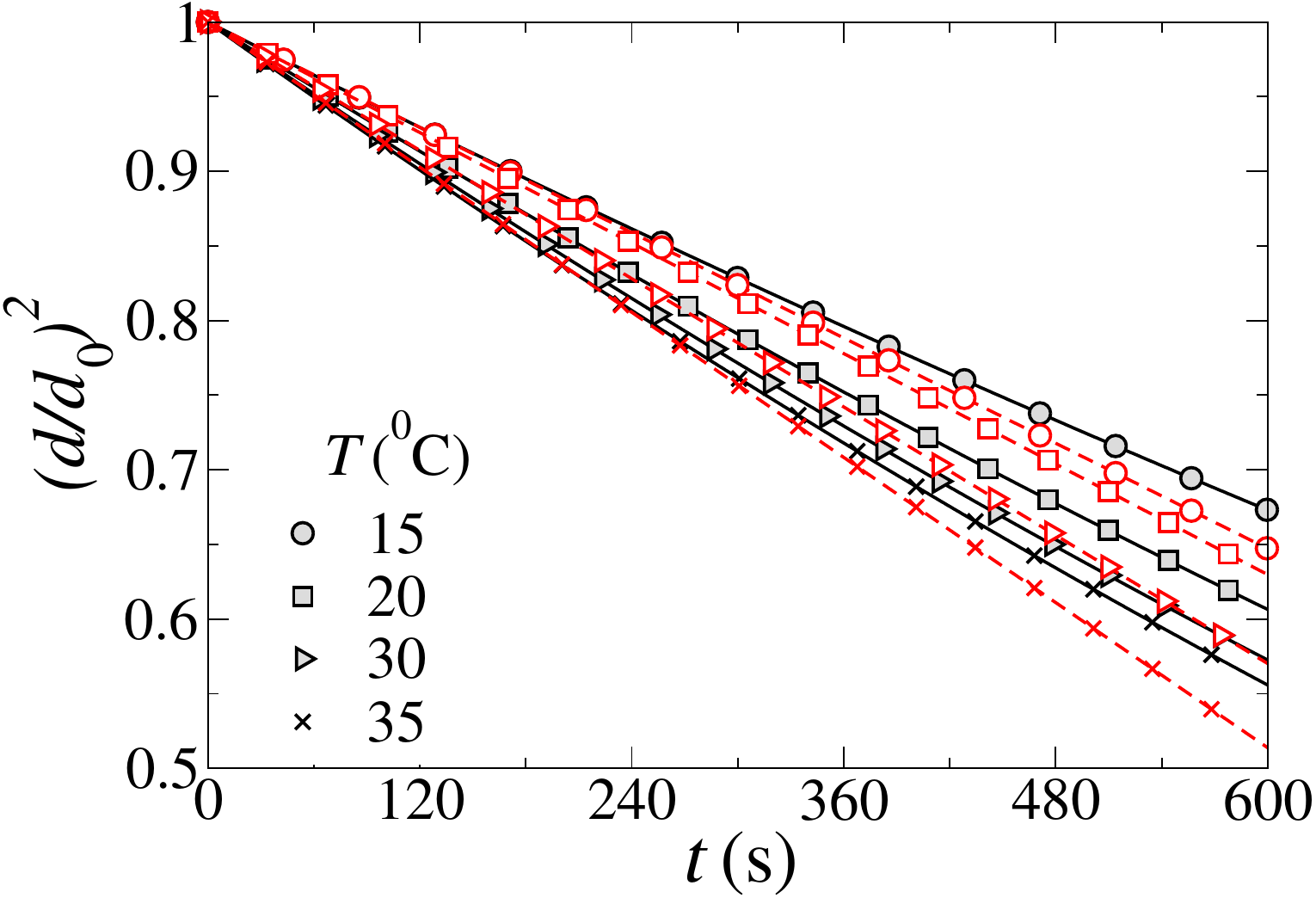} \\
\hspace{0.6 cm} {\large (c)} \\
\includegraphics[width=0.485\textwidth]{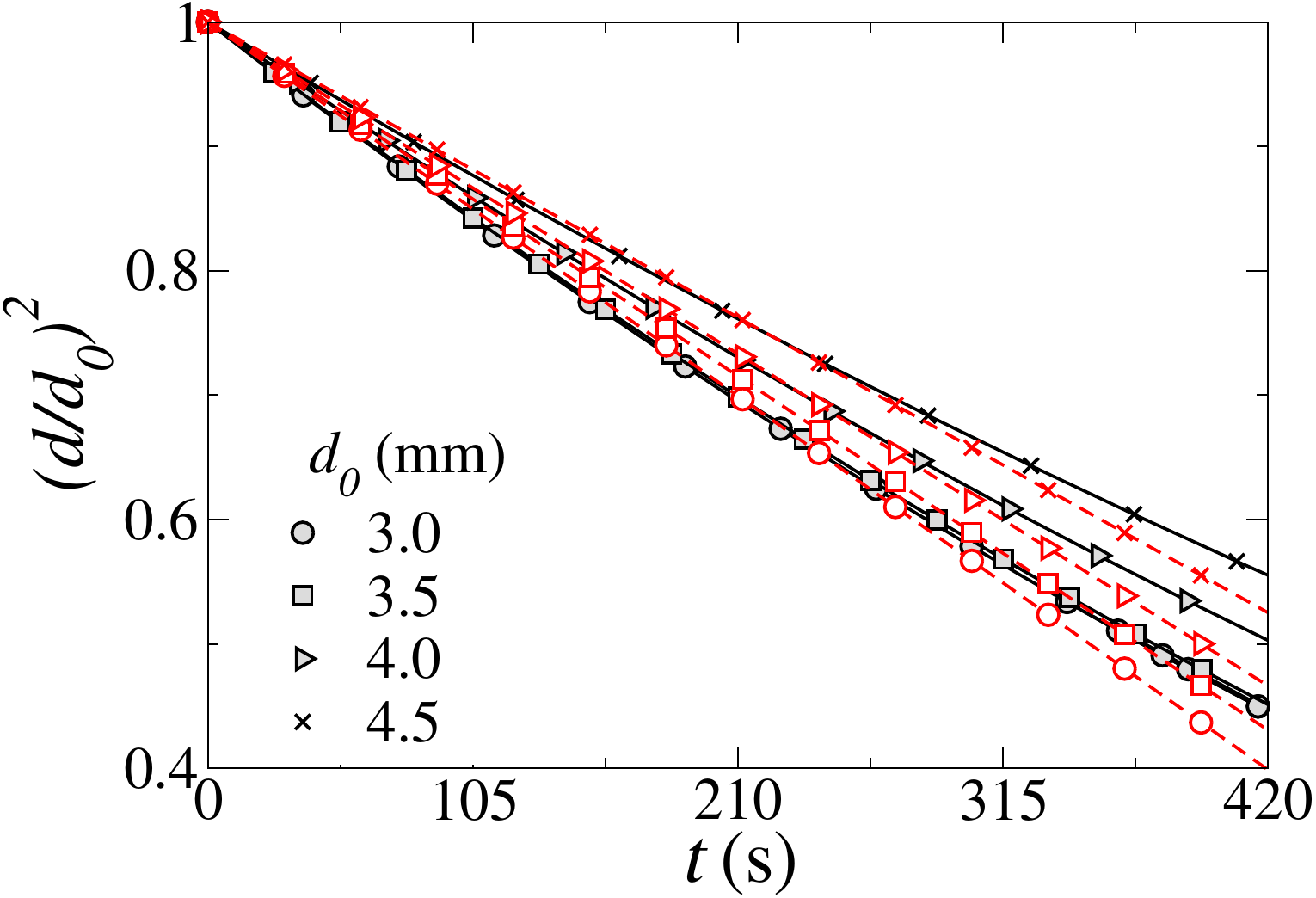}
\caption{Comparison of theoretical predictions based on equation~\eqref{mod_dsquare} with experimental results under different conditions. The temporal variation of $(d/d_0)^2$ for different (a) relative humidity ($RH$) at $T = 30^\circ$C and $d_0 = 3.0 \pm 0.3$ mm; (b) temperature ($T$) at $RH = 50\%$ and $d_0 = 3.0 \pm 0.3$ mm;
(c) initial droplet diameter ($d_0$) at $T = 30^\circ$C and $RH = 10\%$. Here, the solid lines represent the experimental results, while the dashed lines correspond to theoretical predictions.} 
\label{fig:dia_vol_2}
\end{figure}

\begin{figure}
\centering
\hspace{0.6 cm} {\large (a)} \\
\includegraphics[width=0.85\textwidth]{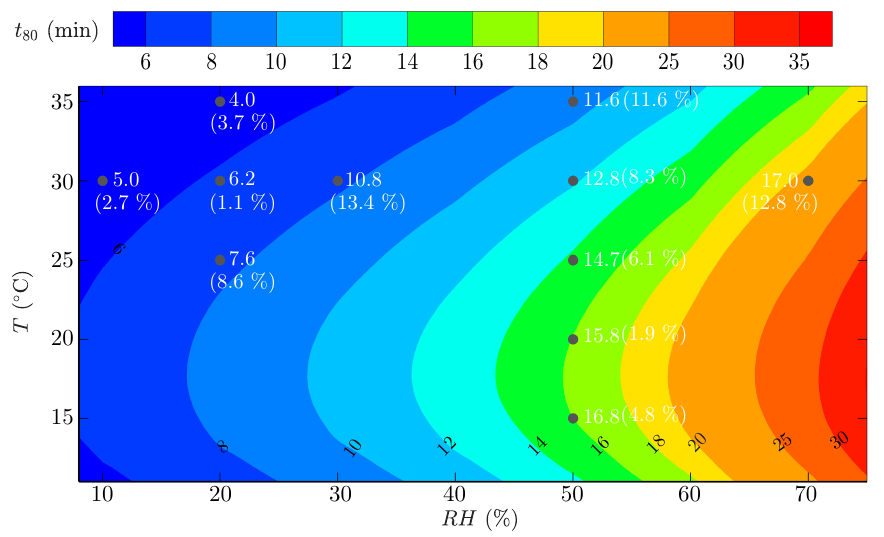}\\
\hspace{0.6 cm} {\large (b)} \\
\includegraphics[width=0.85\textwidth]{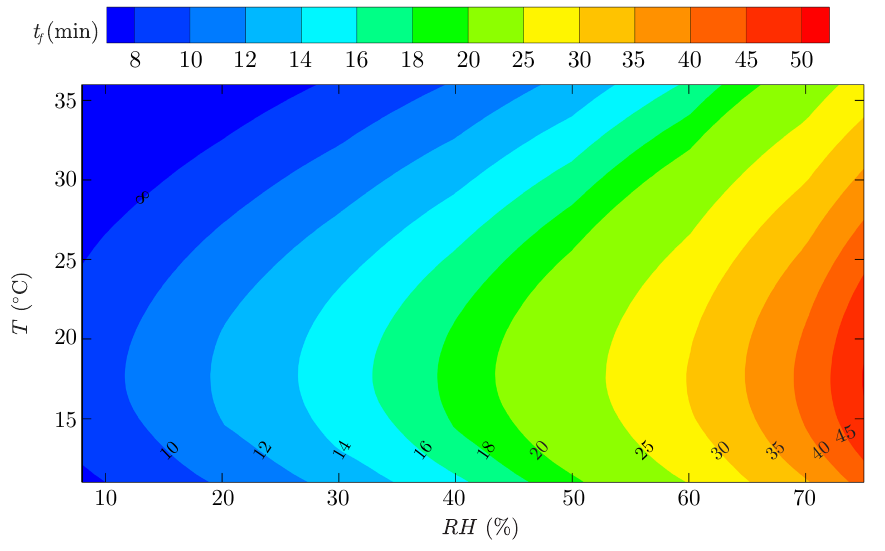}
\caption{Regime map showing the contours of (a) $t_{80}$ and (b) $t_f$ in the relative humidity $(RH)$ – temperature $(T)$ space. The color bars above each panel represent the corresponding values of $t_{80}$ and $t_f$. In panel (a), the filled circles represent the experimental data points, and the percentage deviations between the theoretically predicted and experimentally measured $t_{80}$ values are indicated. The initial droplet diameter is $d_0 = 3.0 \pm 0.3$ mm.}
\label{contour}
\end{figure}

\begin{table}
\centering
\caption{Comparison of the percentage deviation of the droplet diameter, evaluated at 70\% of the initial droplet diameter, as predicted by the classical $d^2$-law (eq. \ref{dsquare}) and the modified $d^2$-law (eq. \ref{mod_dsquare}) with respect to the experimental data for different relative humidity levels ($RH$) at $T = 30^\circ$C and $d_0 = 3.0 \pm 0.3$ mm. The percentage deviation for each case is defined as $(d_\text{experimental} - d_\text{predicted})/d_\text{experimental} \times 100$.}
\label{dev_RH}
\vspace{5mm}
\begin{tabular}{lccc}
& ~~~ RH (\%) ~~~& \shortstack{Percentage deviation  ~~~\\ Classical $d^2$-law (eq. \ref{dsquare})} & \shortstack{Percentage deviation  \\ Modified $d^2$-law (eq. \ref{mod_dsquare})} \\
\
& 10 & 10.95 & 0.71 \\
& 30 & 6.35  & 0.30 \\
& 50 & 3.58  & 0.07 \\
& 70 & 3.26  & 0.50 \\
\end{tabular}
\end{table}

\begin{table}
\centering
\caption{Comparison of the percentage deviation of the droplet diameter, evaluated at 70\% of the initial droplet diameter, as predicted by the classical $d^2$-law (eq. \ref{dsquare}) and the modified $d^2$-law (eq. \ref{mod_dsquare}) with respect to the experimental data, for different temperatures ($T$) at $RH = 50$\% and $d_0 = 3.0 \pm 0.3$ mm. The percentage deviation for each case is defined as $(d_\text{experimental} - d_\text{predicted})/d_\text{experimental} \times 100$.}
\label{dev_T}
\vspace{5mm}
\begin{tabular}{lccc}
 & ~~~ $T$ ($^\circ$C) ~~~& \shortstack{Percentage deviation  ~~~\\ Classical $d^2$-law (eq. \ref{dsquare})} & \shortstack{Percentage deviation  \\ Modified $d^2$-law (eq. \ref{mod_dsquare})} \\
& 15 & 14.40 & 0.68 \\
& 20 & 16.64  & 0.75 \\
& 30 & 15.38  & 0.69 \\
& 35 & 16.14  & 1.8 \\
\end{tabular}
\end{table}

\begin{table}
\centering
\caption{Comparison of the percentage deviation of the droplet diameter, evaluated at 70\% of the initial droplet diameter, as predicted by the classical $d^2$-law (eq. \ref{dsquare}) and the modified $d^2$-law (eq. \ref{mod_dsquare}) with respect to the experimental data, for different initial droplet sizes ($d_0$) at $T = 30^\circ$C and $RH = 10$\%. The percentage deviation for each case is defined as $(d_\text{experimental} - d_\text{predicted})/d_\text{experimental} \times 100$.}
\label{dev_d_0}
\vspace{5mm}
\begin{tabular}{lccc}
 & ~~~ $d_0$ (mm) ~~~& \shortstack{Percentage deviation  ~~~\\ Classical $d^2$-law (eq. \ref{dsquare})} & \shortstack{Percentage deviation  \\ Modified $d^2$-law (eq. \ref{mod_dsquare})} \\
& 3.5 & 9.65 & 0.07 \\
& 3.5 & 11.39  & 1.01 \\
& 4.0 & 11.14  & 0.17 \\
& 4.5 & 9.55  & 0.49 \\
\end{tabular}
\end{table}

\begin{figure}
\centering
\hspace{0.6 cm} {\large (a)} \hspace{6.2 cm} {\large (b)} \\
\includegraphics[width=0.485\textwidth]{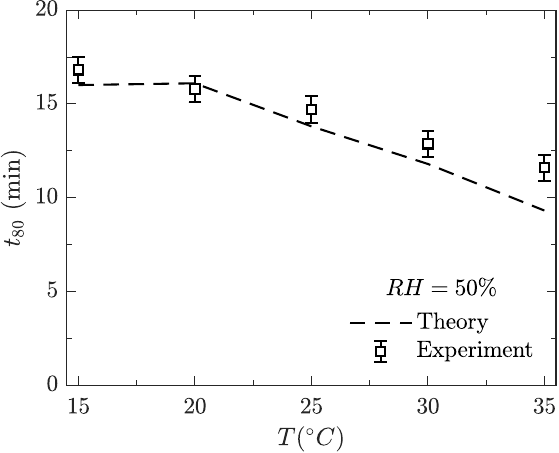} \hspace{0.5mm}
\includegraphics[width=0.485\textwidth]{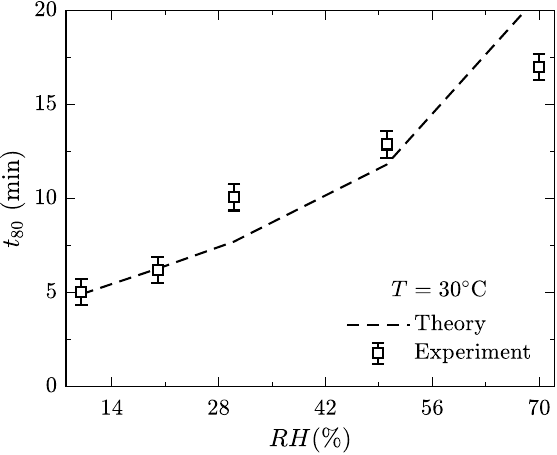} \\
\hspace{0.6 cm} {\large (c)} \\
\includegraphics[width=0.485\textwidth]{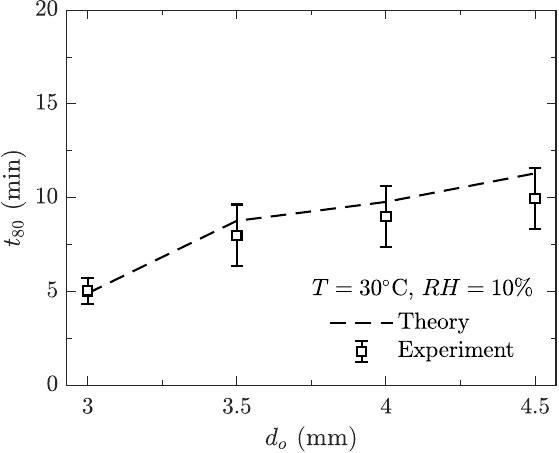}
\caption{Comparison of theoretically predicted values of $t_{80}$, obtained using eq.~\eqref{t80_eva}, with experimental results for different conditions. Variation of $t_{80}$ (in minutes) with (a) temperature ($T$) at $RH = 50\%$ and $d_0 = 3.0 \pm 0.3$ mm; (b) relative humidity ($RH$) at $T = 30^\circ$C and $d_0 = 3.0 \pm 0.3$ mm; (c) initial droplet diameter ($d_0$) at $T = 30^\circ$C and $RH = 10\%$. Error bars at each data point represent the standard deviation of the experimentally measured $t_{80}$ across three repeated trials.}
\label{evapo_time}
\end{figure}

The comparison between experimental results and theoretical predictions based on the modified $d^2$-law (eq.~\ref{mod_dsquare}) for varying relative humidity, temperature, and initial droplet diameter is presented in figures~\ref{fig:dia_vol_2}(a-c), respectively. Here, the solid lines represent the experimental results, while the dashed lines correspond to theoretical predictions. As discussed earlier, during the evaporation process, the square of the droplet diameter (which is proportional to the surface area) decreases linearly with time. This behaviour is consistently observed in both the experiments and the predictions from the modified $d^2$-law. Moreover, the modified theoretical model satisfactorily captures the evaporation rate measured in the experiments. These findings further justify the necessity of including the effects of forced convection and shape oscillations in the evaporation of an aerodynamically suspended droplet, thereby addressing the limitations of the classical $d^2$-law, as discussed in Appendix~\ref{appendix}, to accurately predict the evaporation rate. For completeness, Tables~\ref{dev_RH}, \ref{dev_T}, and \ref{dev_d_0} report the percentage deviation in droplet diameter $(d)$, evaluated at 70\% of the initial diameter, predicted by the classical and modified $d^2$-laws relative to the experimental measurements. These comparisons are presented in Tables~\ref{dev_RH}, \ref{dev_T}, and \ref{dev_d_0} for varying relative humidity ($RH$), temperature ($T$), and initial droplet size ($d_0$), respectively. The percentage deviation for each case is defined as $\left((d_\text{experimental} - d_\text{predicted})/d_\text{experimental} \right) \times 100$. It can be observed from Tables~\ref{dev_RH}, \ref{dev_T}, and \ref{dev_d_0} that while the predictions from the classical evaporation model deviate significantly from the experimental results, the predictions from the modified model show excellent agreement with the experimentally measured droplet sizes, with a maximum deviation of less than 1.8\% and most cases exhibiting deviations below 1\% for the range of parameters considered in this study.

Furthermore, figure~\ref{contour}(a) presents a contour plot of the droplet evaporation time ($t_{80}$) as a function of ambient relative humidity and temperature, computed using eq.~\eqref{t80_eva}. This formulation incorporates the combined effects of liquid properties, diffusion, convective mass transfer, vapour concentration gradients, and droplet shape dynamics. Each color band in the plot represents a specific range of evaporation times. The contour lines denote isolines of constant theoretical evaporation time, with numerical values specified along the lines. To validate the theoretical model, the predicted evaporation times are compared with experimental measurements, shown as grey-filled circles in the plot. The experimentally measured $t_{80}$ values are displayed in white text at the corresponding locations. A close match is observed between the theoretical predictions and the experimental results. The contour plot clearly illustrates that the evaporation time decreases with increasing temperature and decreasing relative humidity. This trend aligns with physical expectations: higher temperatures increase the saturation vapour pressure and enhance diffusion rates, while lower relative humidity intensifies the vapour concentration gradient, contributing to faster evaporation. Eq.~\eqref{t80_eva} effectively captures these dependencies through the Sherwood number ($\Sh$), which accounts for convection; the Spalding mass transfer number ($B$), which reflects vapor gradients; and the average shape factor ($f_{\text{shape,avg}}$), which incorporates the influence of droplet shape oscillations. Collectively, these components provide a comprehensive framework for predicting droplet evaporation under diverse atmospheric conditions.

Finally, the comparison between the experimentally measured evaporation time and the theoretically predicted evaporation time ($t_{80}$) is shown in figures~\ref{evapo_time}(a-c), corresponding to variations in temperature, relative humidity, and initial droplet diameter, respectively. Figure~\ref{evapo_time}(a) illustrates the influence of temperature at a constant relative humidity of $RH = 50\%$ and an initial droplet diameter of $d_0 = 3.0 \pm 0.3$ mm. The theoretical model captures the observed decreasing trend in evaporation time with increasing temperature, which is attributed to enhanced vapour diffusion and higher saturation vapour pressure. Figure~\ref{evapo_time}(b) presents the effect of relative humidity at a fixed temperature of $T = 30^\circ$C and the same initial droplet diameter. The results exhibit good agreement between the theoretical predictions and experimental data, indicating that increased humidity leads to longer evaporation due to reduced vapour concentration gradient. Figure~\ref{evapo_time}(c) examines the role of initial droplet diameter at constant $T = 30^\circ$C and $RH = 10\%$. As expected, the evaporation time increases with droplet size, owing to the larger volume-to-surface-area ratio, which slows the overall evaporation process. Thus, overall, the proposed theoretical model consistently agrees with experimental observations across all cases, confirming its ability to capture the dominant physical mechanisms governing droplet evaporation accurately.

\section{Concluding remarks}\label{sec:conc}
We conduct a combined experimental and theoretical investigation of the evaporation dynamics of freely levitated water droplets in an upward airstream under controlled atmospheric conditions, with an emphasis on simulating raindrop-like behavior. A state-of-the-art wind tunnel facility is employed to precisely control the ambient temperature and relative humidity of the airstream to replicate realistic atmospheric environments. A high-speed imaging system is employed to capture the complex morphology, shape oscillations and temporal evolution of the volume of the droplet during evaporation, thereby enabling detailed characterization of the interplay between aerodynamic forces, surface tension, and environmental conditions. 

Our experimental observations reveal that larger droplets exhibit persistent shape oscillations due to the competition between inertia and surface tension. These oscillations, influenced by airflow and varying temperature and humidity, significantly alter the evaporation rate compared to that of a spherical droplet in a quiescent medium. To quantitatively describe this behavior, we propose a modified evaporation model that extends the classical $d^2$-law by incorporating (i) convective effects through a generalized Sherwood number that accounts for variations in Reynolds number, Schmidt number, temperature, and relative humidity, and (ii) droplet shape oscillations via a time-averaged shape factor. The modified evaporation model shows excellent agreement with experimental results across a wide range of droplet sizes, temperatures, and humidity levels, accurately capturing both the temporal evolution of droplet diameter and the corresponding evaporation time. Furthermore, we construct a regime map showing contours of droplet lifetimes in temperature–humidity space, demonstrating strong agreement between the theoretical predictions and experimental findings across a wide range of environmental conditions. 

To summarize, this study presents a robust and physically consistent framework that captures the effects of convection, complex morphology, and shape oscillations of droplets in an airstream. The results have significant implications for understanding cloud microphysics, weather forecasting, and climate modeling. Future work will extend this framework to include the effects of turbulence, multicomponent droplets, and interactions among multiple droplets in cloud environments.\\

\noindent{\bf Acknowledgement:} We sincerely thank the anonymous reviewers for their insightful comments and constructive suggestions, which significantly improved the quality and depth of the manuscript. \\

\noindent{\bf Funding:} This work was supported by the Suzuki Next Bharat Fellowship Program (Grant No. NBV/CHE/F011/2025-26/S401). \\

\noindent{\bf Declaration of interests:} The authors report no conflict of interest. \\

\noindent{\bf Data availability statement:} Data will be made available on request. \\

\noindent{\bf Author ORCID:}  \\
S. Chakraborty: https://orcid.org/0000-0002-1868-3845 \\
S. S. Ade: https://orcid.org/0009-0007-2652-9358 \\
A. J. Tudu: https://orcid.org/0009-0004-6946-5858 \\
L. D. Chandrala: https://orcid.org/0000-0002-2695-9469 \\
K. C. Sahu: https://orcid.org/0000-0002-7357-1141\\

\noindent{\bf Author contributions:} S.C. established the experimental facility and methodology, conducted the experiments, developed the theoretical model, and wrote the original draft. S. S. A. assisted in setting up the experiments and contributed to the theoretical modelling. A. J. T. supported the experiments and carried out portions of the data plotting and analysis. L. D. C. and K. C. S. contributed to the methodology, investigation, funding acquisition, formal analysis, conceptualization, and preparation of the final manuscript.\\

\appendix

\section{Visualization of the flow-field inside the test section} \label{appendix_PIV}

\begin{figure}
\centering
\includegraphics[width=0.9\textwidth]{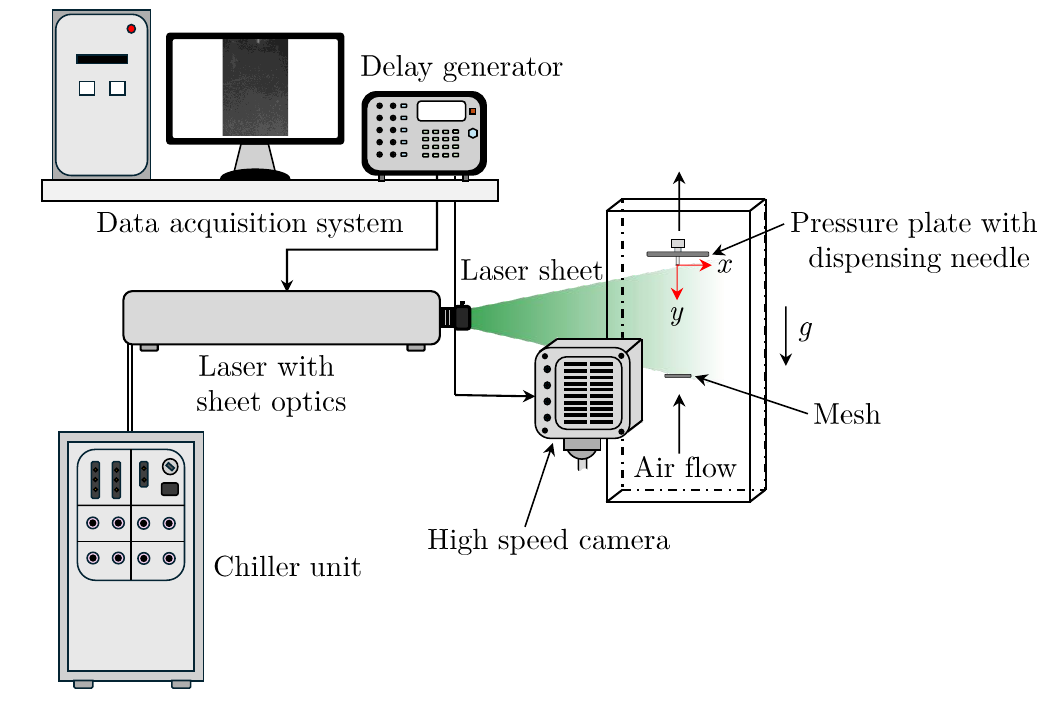}
\caption{Schematic of the Particle Image Velocimetry (PIV) setup used to quantify the flow field inside the test section of the raindrop research facility (figure \ref{RRF}).}
\label{piv_schematic}
\end{figure}

Particle Image Velocimetry (PIV) experiments are performed to visualize and quantify the velocity field within the wind tunnel test section, which features a pressure plate at the top and a mesh at the bottom. A schematic of the PIV setup is shown in figure \ref{piv_schematic}. The system comprises: (i) a double-pulsed Nd:YAG laser (Litron Nano L200-15-LM3128, $\lambda$=532 nm), (ii) sheet-forming optics consisting of spherical and cylindrical lenses, (iii) a high-speed camera (Phantom VEO 640L, Vision Research, USA) equipped with a 135~mm focal-length lens (Nikkor, $f$/2 minimum aperture), (iv) a digital delay generator (610036, TSI, USA), and (v) a data acquisition computer. The sheet-forming optics converts the laser beam into a planar light sheet approximately 0.5~mm thick, which illuminates the test section midplane. The flow is seeded with oil-based tracer particles (1--5~$\mu$m diameter) generated by Laskin nozzle. The high-speed camera is positioned perpendicular to the laser sheet, providing a field of view of approximately 143~mm~$\times$~98~mm (1500~pixels~$\times$~1024~pixels). To achieve a particle displacement of approximately 25\% within the interrogation window, the inter-pulse delay between successive laser pulses is set to 50~$\mu$s. Image pairs are acquired at 10~Hz, with the digital delay generator synchronizing the laser pulse timing and camera exposure sequence.

Next, the recorded images are processed following the methodology described in our previous work \citep{chakraborty2025drop}, using PIVlab, which is an open-source \textsc{MATLAB}$^{\circledR}$ based graphical interface \citep{thielicke2021particle}. The post-processing using this software involves several steps to extract the velocity field from the raw images. The workflow begins with Fast Fourier Transform (FFT) based cross-correlation to compute displacement vectors between successive frames. To improve spatial resolution and enhance the signal-to-noise ratio, multi-pass refinement and multi-grid window deformation algorithms are employed. The first interrogation pass uses a $128 \times 128$ pixel window with 50\% overlap (step size: 64 pixels), followed by a refinement pass using a $64 \times 64$ pixel window (step size: 64 pixels). The sub-pixel displacement estimation employs a Gaussian peak-fitting method. Finally, the velocity field is obtained by averaging 200 image pairs to suppress temporal fluctuations and yield a representative flow distribution within the test section.

\begin{figure}
\centering
\hspace{0.6 cm} {\large (a)} \hspace{6.2 cm} {\large (b)} \\
\includegraphics[width=0.485\textwidth]{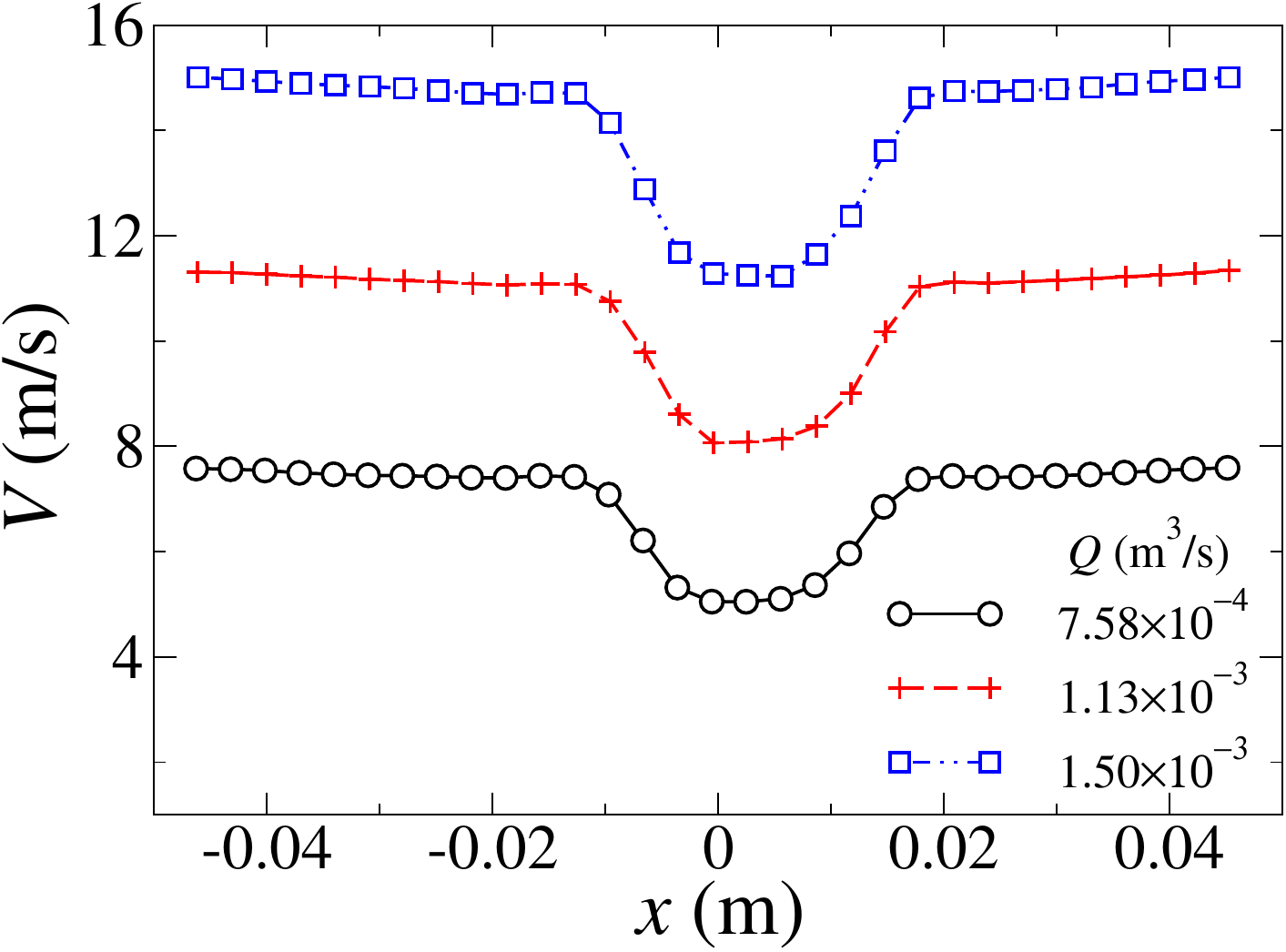} \hspace{0.5mm}
\includegraphics[width=0.485\textwidth]{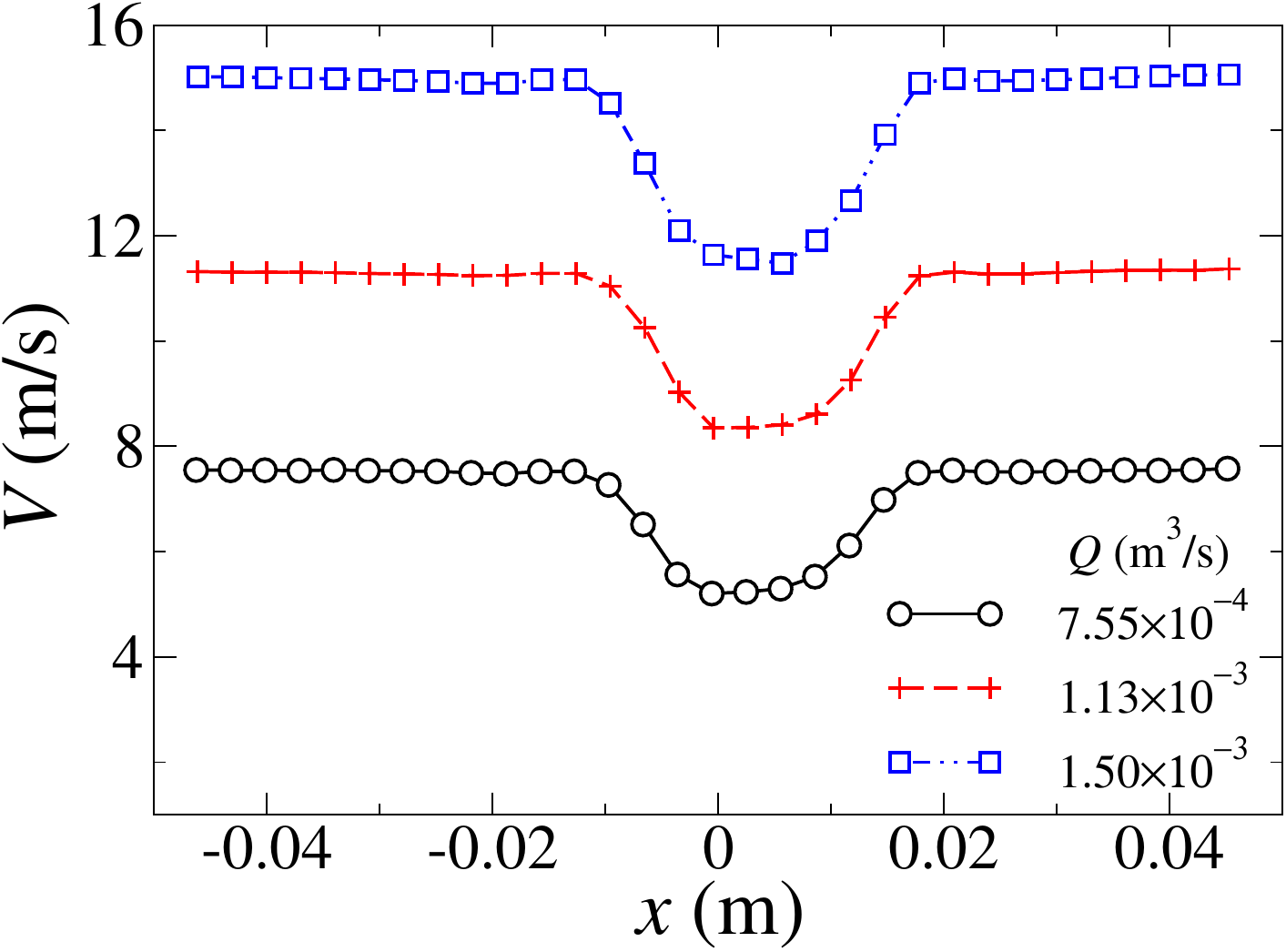} \\
\hspace{0.6 cm} {\large (c)}  \\
\includegraphics[width=0.485\textwidth]{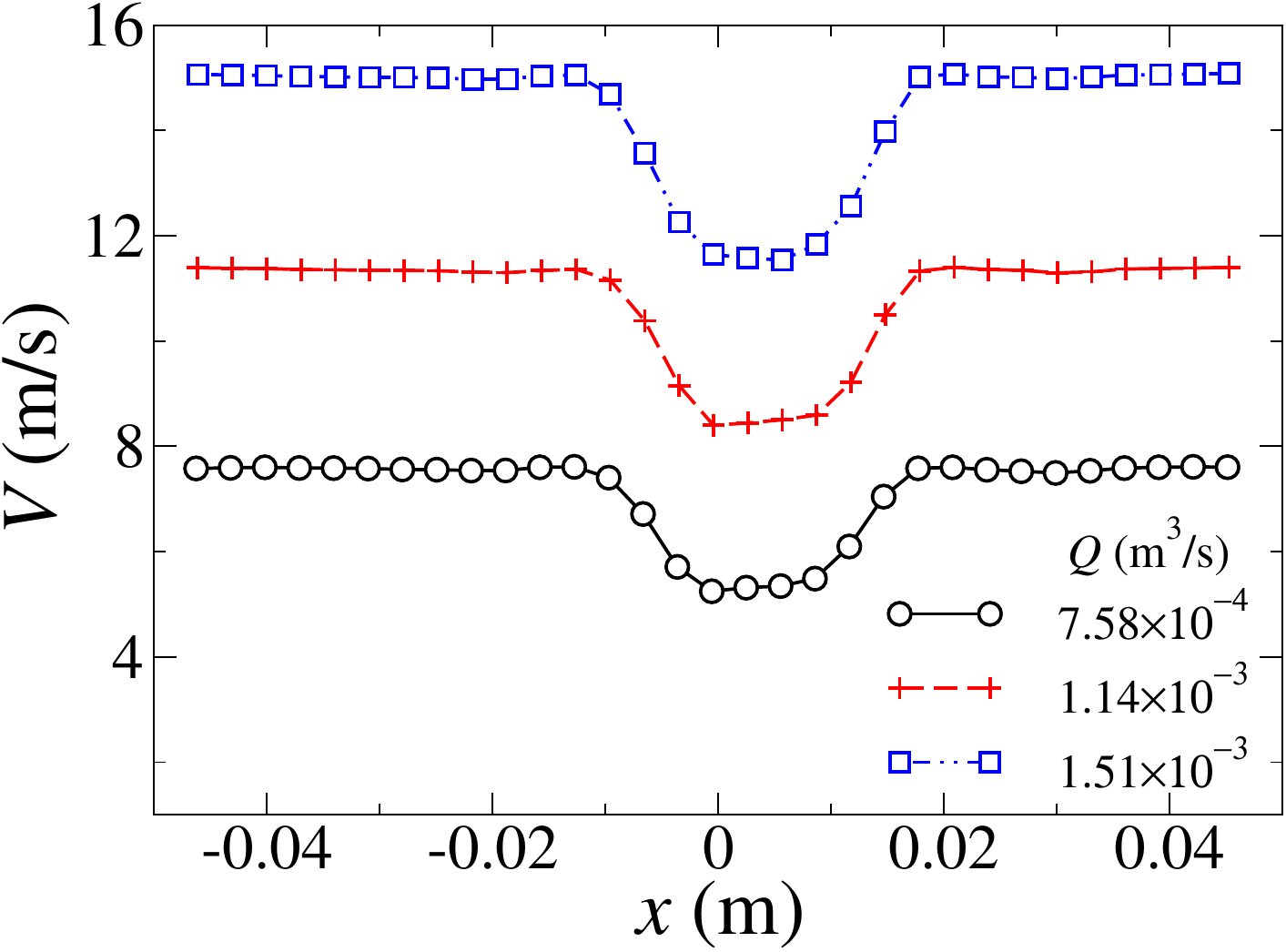}
\caption{Velocity profiles obtained from Particle Image Velocimetry (PIV) measurements at three vertical locations for different flow rates: at (a) $y = 5$ cm, (b) $y = 7$ cm, and (c) $y = 9$ cm from the tip of the dispensing needle.}
\label{vel_prof}
\end{figure}

The measured velocity profiles for different flow rates at three vertical locations, namely at $y = 5$ cm, $y = 7$ cm, and $y = 9$ cm from the tip of the dispensing needle, are shown in figure \ref{vel_prof}(a–c), respectively. It can be seen that the velocity distribution remains nearly uniform away from the centerline of the test section. However, a distinct dip, referred to as the velocity well, is observed near the center region of the test section. This feature arises due to the presence of the mesh plate at the bottom (see figure \ref{piv_schematic}), while the pressure plate at the top generates a back pressure that aids in droplet dispensing. The velocity well plays a key role in stabilizing the droplet by preventing it from being transported with the airflow. The velocity well acts as a hydrodynamic trap, where the pressure maximum at the center (where the velocity is minimum) creates a stable equilibrium position. Any lateral displacement generates a pressure-gradient restoring force that drives the drop back to center, analogous to a ball settling at the bottom of a mechanical potential well. The depth of the velocity well remains nearly unchanged across the vertical positions shown in figure \ref{vel_prof}(a–c), while, as expected, the average velocity increases with increasing flow rate ($Q$). A similar velocity well was previously reported by \citet{Kamra1991}, who investigated the effect of an electric field on freely suspended water droplets in a wind tunnel.

\section{Classical evaporation model} \label{appendix}

\begin{figure}
\centering
\includegraphics[width=0.6\textwidth]{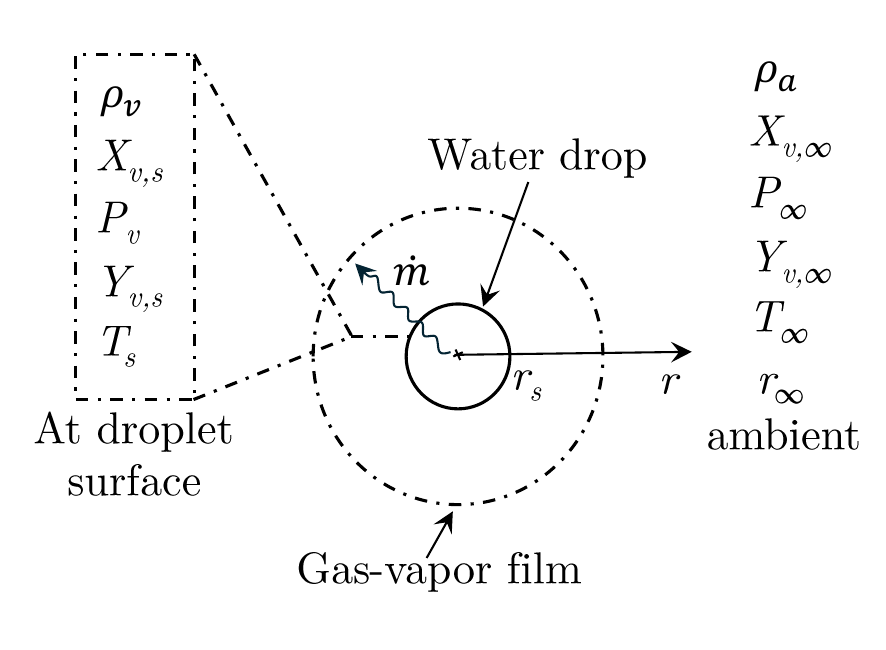}
\caption{Schematic representation of the evaporation mechanism for a droplet surrounded by a gas-vapor film in a gaseous medium. Here, $\dot{m}$ denotes the mass flow rate from the droplet surface into the surrounding gas-vapor region along the radial direction ($r$). The properties and parameters at the surface of the droplet ($r = r_{s}$) and at the ambient ($r \rightarrow r_{\infty}$) have been illustrated.}
\label{drop_basic}
\end{figure}

The rate of evaporation of a water droplet in air is primarily governed by the transport of water vapour through a combination of advection and diffusion mechanisms. Figure~\ref{drop_basic} depicts a schematic of the evaporation process for a droplet surrounded by a gas–vapour film in a gaseous medium. The solid boundary denotes the droplet surface, while the surrounding dashed region represents the gas–vapour film. The region beyond this film corresponds to the ambient environment. Physical properties vary along the radial direction $r$, and the rate of evaporation, denoted by $\dot{m}$, is illustrated by a wavy arrow directed outward from the droplet surface into the gas–vapour region.

Assuming that (i) the process is quasi-steady, (ii) the temperature within the droplet is spatially uniform, (iii) the vapor mass fraction at the droplet surface equals its saturation value, and (iv) physical properties such as the saturated vapor density and binary diffusivity remain constant, the mass flux from the droplet surface in a spherically symmetric coordinate system can be described by Fick’s law is given by \citep{law1982recent}:
\begin{equation}
\dot{m} = \dot{m} {Y_{v}} - 4\pi r^2 {D_{v}} {\rho_v} \left( \frac{{dY_{v}}}{dr} \right), \label{mass}
\end{equation}
where $Y_{v}$ is the mass fraction of water vapor, $\rho_v$ is the saturated water vapor density, $D_{v}$ is the binary diffusion coefficient, $r$ is the radial coordinate, and ${dY_{v}}/dr$ is the radial gradient of vapor mass fraction.

This is a first-order ordinary differential equation, which can be solved using the boundary condition at the droplet surface:
\begin{equation}
{Y_{v} = Y_{v,s}} ~{\rm at}~ r = r_s
\end{equation}
to obtain the expression for the vapour mass fraction, which is given by
\begin{equation}
{Y_{v}} = 1 - \frac{(1 - {Y_{v,s}}) \exp\left(-\frac{\dot{m}}{4\pi {\rho_v} {D_{v}} r} \right)}{\exp\left(-\frac{\dot{m}}{4\pi {\rho_v} {D_{v}} r_s} \right)}. \label{Yw_eqn}
\end{equation}
Then, imposing the far-field boundary condition,
\begin{equation}
{Y_{v}} = {Y_{v,\infty}} ~ {\rm at} ~ r \rightarrow r_{\infty},
\end{equation}
and substituting into eq.~\eqref{Yw_eqn}, we obtain the expression for the evaporation rate \citep{sirignano2010droplets}:
\begin{equation}
\dot{m} = 4\pi r_s {D_{v}} {\rho_v} \ln\left( \frac{1 - {Y_{v,\infty}}}{1 - {Y_{v,s}}} \right). \label{mod_eva}
\end{equation}
Here, $Y_{v,s}$ and $Y_{v,\infty}$ are the mass fractions of water vapour at the droplet surface (saturation) and in the ambient environment, respectively. These can be calculated using the following relationships \citep{pinheiro2019evaluation}:
\begin{eqnarray}
{Y_{v,s}} = \frac{{X_{v,s}} {M_v}}{{X_{v,s} M_v} + (1 - {X_{v,s}}) {M_{a}}}, \label{yvs} \\
{Y_{v,\infty}} = \frac{{X_{v,\infty}}{M_v}}{{X_{v,\infty} M_v} + (1 - {X_{v,\infty}}) {M_{a}}}, \label{yvinfi}
\end{eqnarray}
where $X_{v,s}$ is the mole fraction of water vapour at the droplet surface, and $X_{v,\infty}$ is the mole fraction in the ambient environment. According to Raoult’s law, $X_{v,s}$ is given by:
\begin{equation}
{X_{v,s}} = \frac{{P_v}}{P_{\infty}}, \label{xv,s}
\end{equation}
while $X_{v,\infty}$ can be written as:
\begin{equation}
{X_{v,\infty}} = RH \left( \frac{{P_v}}{P_{\infty}} \right), \label{xinfi}
\end{equation}
where $P_v$ is the saturation vapor pressure of water and $P_{\infty}$ is the total ambient pressure. $RH$ denotes the relative humidity. The molecular weights are ${M_v} = 18.01$ g/mol for water vapour and $M_{a}$ = $28.97$ g/mol for dry air. Eq. ~\eqref{mod_eva} can also be expressed in terms of the instantaneous droplet mass ($m_d$) and diameter ($d = 2r_s$) as:
\begin{equation}
\frac{d m_d}{dt} = \frac{d}{dt} \left( \rho_w \frac{\pi d^3}{6} \right) = -2\pi d {D_{v} \rho_v} \ln\left( \frac{1 - {Y_{v,\infty}}}{1 - {Y_{v,s}}} \right). 
\label{drop_mass}
\end{equation}
Solving this equation yields the evaporation rate constant $K$ as:
\begin{equation}
K = \frac{8 {\rho_v D_{v}} \ln(1 + B)}{\rho_w}, \label{k}
\end{equation}
with $B$ being the Spalding mass transfer number, defined as \citep{spalding1960standard}:
\begin{equation}
B = \frac{{Y_{v,s}} - {Y_{v,\infty}}}{{1 - Y_{v,s}}}. \label{mass_tran_number}
\end{equation}
This leads to the classical $d^2$-law for droplet evaporation \citep{langmuir1918evaporation,law1982recent}:
\begin{equation}
d^2(t) = d_0^2 - K t, \label{dsquare}
\end{equation}
where $d_0$ is the initial droplet diameter at $t = 0$.

\begin{figure}
\centering
\hspace{0.6 cm} {\large (a)} \hspace{6.2 cm} {\large (b)} \\
\includegraphics[width=0.485\textwidth]{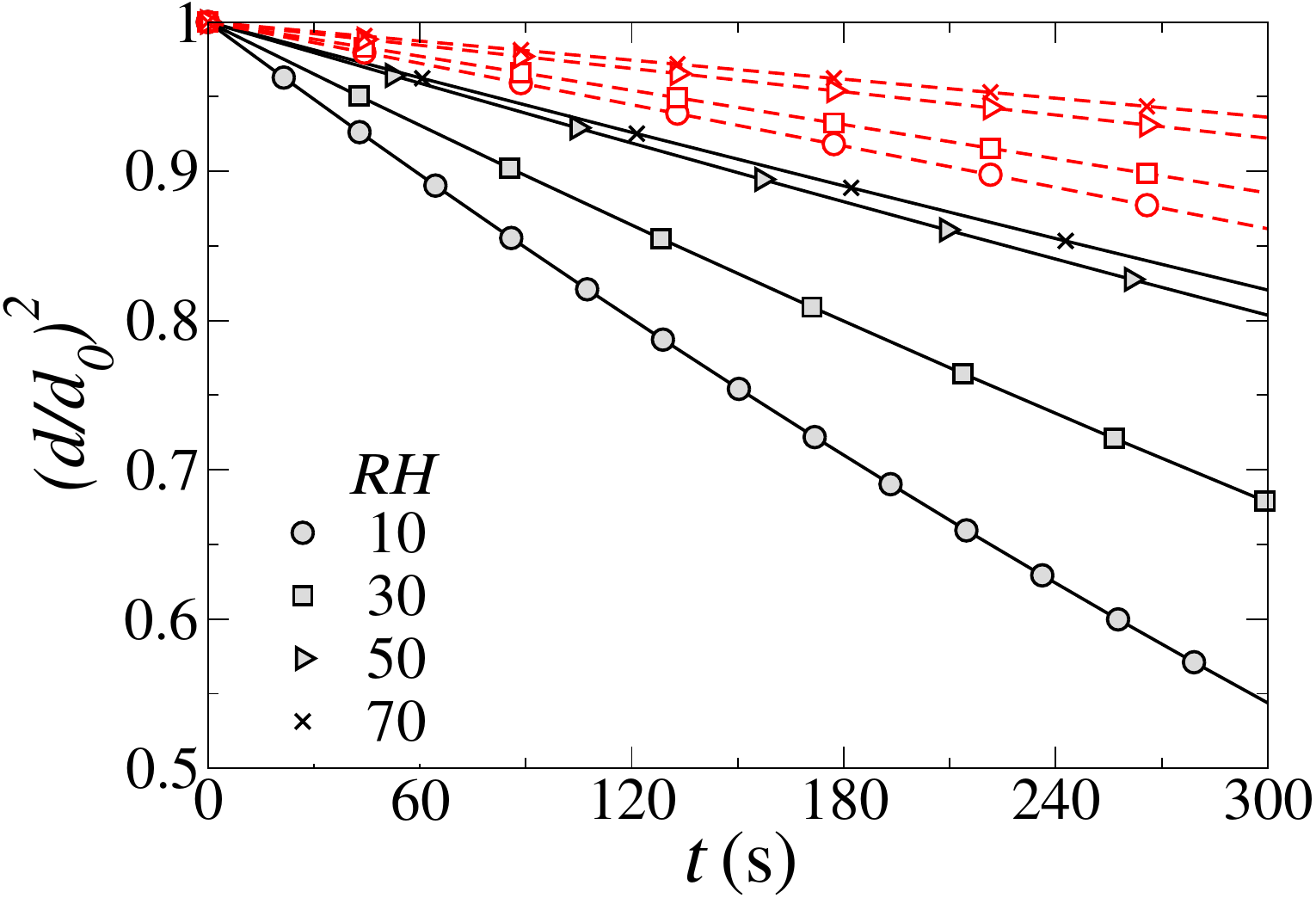} \hspace{0.5mm}
\includegraphics[width=0.485\textwidth]{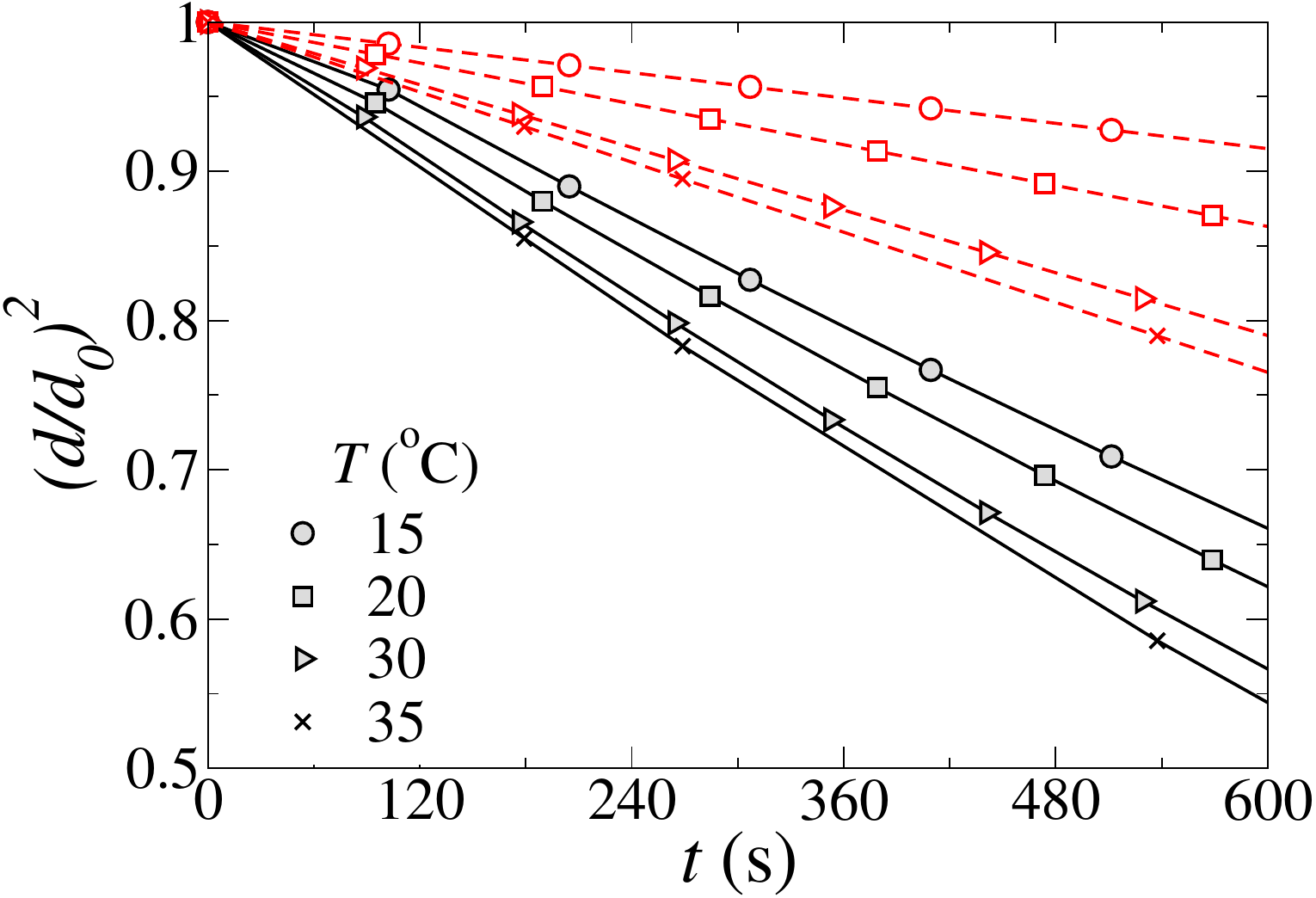} \\
\hspace{0.6 cm} {\large (c)}  \\
\includegraphics[width=0.485\textwidth]{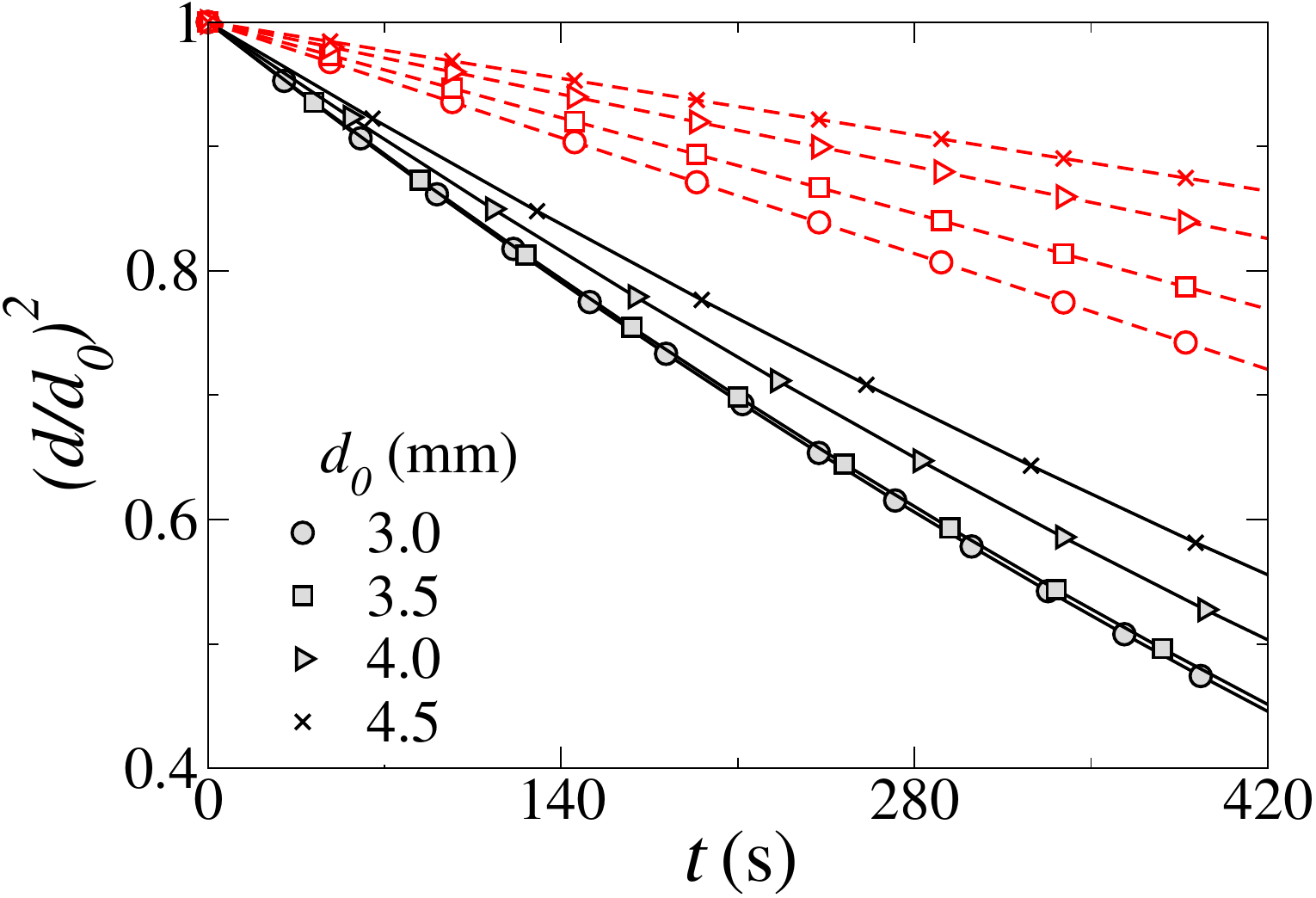}
\caption{Comparison of theoretical predictions obtained using eq.~\eqref{dsquare} with the experimental results for different conditions. The temporal variation of $(d/d_0)^2$ for different (a) relative humidity ($RH$) at $T = 30^\circ$C and $d_0 = 3.0 \pm 0.3$ mm; (b) temperature ($T$) at $RH = 50\%$ and $d_0 = 3.0 \pm 0.3$ mm;
(c) initial droplet diameter ($d_0$) at $T = 30^\circ$C and $RH = 10\%$. Here, the solid lines represent the experimental results, while the dashed lines correspond to theoretical predictions.}
\label{fig:dia_vol_4}
\end{figure}

Figures~\ref{fig:dia_vol_4}(a-c) present the comparison between experimental results and theoretical predictions based on eq.~\eqref{dsquare} for different values of the relative humidity $(RH)$, temperature $(T)$, and initial droplet diameter $(d_0)$. The solid lines in figures~\ref{fig:dia_vol_4}(a–c) represent the experimental data, while the dashed lines correspond to the theoretical predictions based on the $d^2$-law. As discussed earlier, an increase in relative humidity ($RH$) results in a lower evaporation rate, whereas higher temperatures ($T$) enhance the evaporation rate. Likewise, the evaporation rate decreases with increasing initial droplet diameter ($d_0$), due to the reduced surface-area-to-volume ratio. In figure~\ref{fig:dia_vol_4} (a-c), it is evident that the square of the droplet diameter, which is proportional to the surface area, decreases linearly with time throughout the evaporation process. This linear trend is consistent with both the experimental observations and the theoretical prediction from the $d^2$-law. However, the theoretical model tends to significantly underpredict the evaporation rate compared to the experimental data. This discrepancy arises because the evaporation rate constant $K$ in the $d^2$-law depends on the Spalding mass transfer number ($B$), which is a function of the mass fraction ratio at the droplet surface ($Y_{v,s}$) and in the ambient environment ($Y_{v,\infty}$). Since this ratio appears inside a natural logarithm in the expression for $K$, even relatively large values of $B$ exert a limited influence on the overall evaporation rate. As a result, the theoretical model systematically underestimates the actual evaporation rate. We observe that the diffusivity of water vapour in air plays the most dominant role in determining the evaporation rate and droplet lifetime, particularly when $B$ is small. In such cases, the decay of $d_0^2$ is governed primarily by vapour diffusivity, rather than by the difference in mass fractions alone. In summary, while the $d^2$-law offers a valuable framework for understanding droplet evaporation, it remains a diffusion-limited model and is most accurate when diffusion is the dominant mass transport mechanism.


\end{document}